\documentclass[usenatbib]{pasa}%
\usepackage{graphicx}

\def \msun{\ifmmode{{\rm\ M}_\odot}\else{${\rm\ M}_\odot$}\fi}

\def \lsun{\ifmmode{{\rm\ L}_\odot}\else{${\rm\ L}_\odot $}\fi}

\def \msun{\ifmmode{{\rm\ M}_\odot}\else{${\rm\ M}_\odot$}\fi}

\def \rsun{\ifmmode{{\rm\ R}_\odot}\else{${\rm\ R}_\odot$}\fi}

\def \mdot{\ifmmode{{\rm\dot{M}}}\else{${\rm\dot{M}}$}\fi}

\newcommand\am{${'}$}

\newcommand\as{${''}$}

\newcommand{\ha}{H$\alpha${}}

\newcommand{\hb}{H$\beta${}}

\newcommand{\hii}{H\,{\sc ii}{}}

\newcommand{\oii}{[O\,{\sc ii}]}

\newcommand{\oiii}{[O\,{\sc iii}]}

\newcommand{\nii}{[N\,{\sc ii}]}

\title[SN environments]{Statistical studies of supernova environments}
\author[Anderson et al.]{Joseph P. Anderson$^1$, Phil A. James$^2$, Stacey M. Habergham$^2$,
Llu\'is Galbany$^{3,4}$ \and Hanindyo Kuncarayakti$^{3,4}$
\\
\affil{$^1$European Southern Observatory, Alonso de Cordova 3107, Vitacura, Casilla 19001, Santiago, Chile}%
\affil{$^2$Astrophysics Research Institute,
Liverpool John Moores University, IC2, Liverpool Science Park, 146 Brownlow Hill,\\
Liverpool, L3 5RF, UK}
\affil{$^3$Millennium Institute of Astrophysics, Santiago, Chile}
\affil{$^4$Departamento de Astronom\'ia, Universidad de Chile, Santiago, Casilla 36-D, Chile}
}%
\jid{PASA}
\doi{10.1017/pas.\the\year.xxx}
\jyear{\the\year}

\usepackage[authoryear]{natbib}
\bibpunct{(}{)}{;}{a}{}{,}
\begin{document}%
\begin{abstract}
Mapping the diversity of supernovae (SNe) to progenitor properties is key to our 
understanding of stellar evolution and
explosive stellar death. Investigations of the immediate environments
of SNe allow statistical constraints to be made on 
progenitor properties such as mass and metallicity.
Here we review the progress that has been made in this field.
Pixel statistics using tracers of e.g. star formation 
within galaxies show intriguing differences in the 
explosion sites of, in particular SNe types II and Ibc (SNe~II and SNe~Ibc respectively), 
suggesting statistical differences in population ages. Of particular interest is
that SNe~Ic are significantly more associated with host galaxy \ha\ emission than SNe~Ib, implying
shorter lifetimes for the former.
In addition, such studies
have shown (unexpectedly) that the interacting SNe~IIn do not explode in regions
containing the most massive stars, which suggests that at least a significant fraction
of their progenitors arise from the lower end of the core-collapse SN mass range.
Host \hii\ region spectroscopy has been obtained for a significant
number of core-collapse events, however definitive conclusions
on differences between distinct SN types have to-date 
been elusive. Single stellar evolution models predict that the relative fraction
of SNe~Ibc to SNe II should increase with increasing metallicity, due to the dependence
of mass-loss rates on progenitor metallicity. 
We present a meta-analysis of all current host \hii\ region oxygen abundances for CC SNe.
It is concluded that the SN~II to SN~Ibc ratio shows little variation with
oxygen abundance, with only a suggestion that the ratio increases in the lowest bin.
Radial distributions of different SNe are 
discussed, where a central excess of SNe~Ibc has been observed within disturbed galaxy systems, which 
is difficult to ascribe to metallicity or selection effects.
Environment studies are also being undertaken for SNe~Ia, where constraints can be made on
the shortest delay times of progenitor systems. It is shown that `redder' SNe~Ia
are more often found within star-forming regions.
Environment studies are evolving to enable studies at higher spatial resolutions than previously possible, while 
in addition the advent of wide-field integral field unit instruments allows galaxy-wide spectral analyses which will provide
fruitful results to this field. Some example contemporary results are shown in that direction.
\end{abstract}
\maketitle%
\section{INTRODUCTION}
Supernovae (SNe) are the explosive deaths of stars. The light released is such that 
the explosion can for a brief time of weeks to months, be
close in luminosity to the total light produced by its host galaxy, i.e. billions of individual stars. 
These extreme luminosities and their associated energies mean that SNe are not only interesting
to understand in their own right, but are also important objects which heavily influence their 
immediate environments and the evolution of those regions. In addition, SNe can also be used
as probes to aid in our understanding of many other astrophysical questions.\\
\indent There now exists a large diversity of SN type classifications, with new distinct events being 
added to the traditional observational classes. 
SNe were originally classified into types I and II,
due to the absence in the former and the presence in the latter, of hydrogen in their spectra \citep{min41}.
Later, the type I class was further sub-divided as it became clear that further distinct types exist.
SNe~Ib show helium in their spectra, while SNe~Ic do not, and both lack the strong silicon
absorption characteristic of SNe~Ia (see \citealt{fil97} 
for a review of SN spectral classifications)\footnote{Throughout the manuscript we will refer
to the combined sample of SNe~Ib, SNe~Ic and those classified as `Ib/c', as SNe~Ibc.}.
SNe~II can also be further split by properties of their light-curve and spectra. SNe~IIP are
observed to have a `plateau' in their light-curves, while the faster declining `linear' events
are classified as SNe~IIL (\citealt{bar79}, although see \citealt{arc12}, \citealt{and14a} and \citealt{far14b} for discussion
on SN~II typing). Two further type II spectroscopic classifications exist. SNe~IIn are those events which
show narrow emission features in their spectra \citep{sch90}, indicating that their ejecta are
interacting
with slow moving circumstellar material (CSM, \citealt{chu94}). At early times SNe~IIb
show features similar to normal SNe~II, but later hydrogen features
disappear, and they appear more similar to SNe~Ib \citep{fil93}. In addition, there
also exist a subset of the SN~Ic class which show broad spectral features implying
high ejecta velocities. These are known as `broad-line SNe~Ic' (or SNe~Ic BL), and a further
subset of SNe~Ic BL have been linked to long duration Gamma Ray Bursts (LGRBs, see e.g. \citealt{gal98,hjo03}).\\
\indent From a theoretical viewpoint, historically SN progenitors have been separated
into two classes. SNe~Ia are presumed to arise from the thermonuclear disruption of a white dwarf (WD)
or WD system, while CC SNe (i.e. types II and Ibc) are thought to be the result of the core-collapse
of massive stars. This separation is given further credence by the fact that CC SNe are 
only observed in star-forming late-type galaxies, while SNe~Ia are observed in all
galaxy types (including ellipticals, \citealt{van05}). Given that elliptical galaxies are dominated by
evolved stellar populations, this effectively constrains the SNe occurring within those
galaxies to arise from similarly evolved progenitors. A WD progenitor can easily accommodate
this constraint. Meanwhile, the fact that CC SNe are exclusively found within galaxies containing
significant star formation (SF), is consistent with their progenitors
being
the collapse of the cores of massive stars. However, outside of these very broad progenitor
constraints of young (CC SNe) and old (SNe~Ia) parent populations, gaining
further information on the specific progenitor parameter space which defines the
observed transient diversity has been relatively slow going.\\
\indent In recent years the traditional classification scheme has been further complicated by the addition of 
peculiar types such as SNe~Iax (see e.g. \citealt{fol13} and references therein) and `Ca-rich' events
(e.g. \citealt{per10}). The features in these distinct transients give clues as to the 
initial progenitor and pre-SN characteristics giving rise to the observed SN diversity. Indeed,
there is debate as to how these events can be understood in terms of 
the classic supposedly fundamental theoretical classes.\\
\indent It is generally accepted that CC SN diversity can be explained through the amount of the progenitor star's envelope retained at
the epoch of explosion. This then defines the dominant spectral features, and 
the shape of the light-curve. The main
question then becomes: which progenitor property defines the amount of mass lost by the star or
stellar system before explosion? Initial progenitor mass, metallicity, and rotation, together
with the presence and influence of a binary companion all affect this degree of mass loss.
The epoch and rate of that mass loss can also heavily influence the observed SN, where in e.g.
SNe~IIn, observations suggest that mass-loss occurred close to the epoch of explosion, 
causing the large dispersion in properties observed in this class (see e.g. \citealt{kie12,tad13}).
Deciphering which changes in progenitor properties link to changes in SN features
is then a key question. Studying the immediate environments where SNe explode can help elucidate these issues.\\
\indent The most direct way to constrain SN progenitors is to observe the progenitor
star before explosion. This has been possible in a small number of cases, and 
for SNe~IIP these studies have constrained their progenitors to be red supergiant (RSG)
stars with initial masses between 8--16\msun\ (see \citealt{sma09b} for a review).
This technique requires that pre-SN images of sufficient depth at SN positions are available.
One can then use photometry of the star coincident 
with explosion locations to derive a luminosity and convert 
this to a progenitor mass using the predictions of stellar evolution codes. While this
technique is highly enlightening with respect to individual SN events, its use in a 
statistical sense is limited by a) the lack of 
explosions close enough where one can detect/resolve individual stars, and b) the need
for sufficiently deep pre-explosion images. Hence, to build up statistics for the 
different SN types through this avenue will take several more decades. At the other 
end of the progenitor constraint spectrum is the analysis of global host galaxy
properties. Indeed as mentioned above, the separation of SNe into CC and thermonuclear events owes
much to the appearance of the latter in early-type elliptical galaxies.\\
\indent More recent studies have investigated how e.g. the SN~Ibc to SN~II ratio changes with
host galaxy mass/luminosity \citep{pra03,boi09,arc10}, metallicity \citep{pri08b}, 
host morphology \citep{hak14}, and the multiplicity of SNe within different galaxies \citep{and13}. 
These studies show intriguing differences, and allow for
statistically significant analyses.
However, especially in late-type
spiral galaxies, there are multiple stellar populations of distinct ages, metallicities and
possible SF scenarios. This therefore makes drawing definitive conclusions on progenitor
properties complicated. An interesting avenue for investigation which overcomes some 
of the limitations of the above is to analyse the stellar populations at the locations 
of SNe within host galaxies. In this direction one can still obtain significant statistics (as
will be shown below), while analysing stellar populations that are more representative 
of SN progenitors than global host investigations.\\
\indent In this review we will summarise the progress that has been made in the field of SN `environments'.
The review is outlined as follows. First, we continue this introduction with 
a short history of environmental studies, discussing the important constraints that have been made
on progenitors, and bringing the 
reader up-to-date with
the previous decade. The introduction finishes
with a description of the definition of what we mean when analysing SN `environments'.
We then outline the different statistical techniques that have recently been used in this field and
this is followed by a discussion of the important results derived from environment analysis in
\S\ 3. In \S\ 4 we further discuss the implications of these results, and the various biases and 
complications in their interpretation. In \S\ 5 the current and future status of this field is 
discussed, and finally in \S\ 6 we conclude with an overall summary.

\subsection{A short history of supernova environment studies}
The first environmental evidence used to constrain the physical
properties of SNe arose due to the observation that type I SNe (at that
stage no separation was made between type Ia and type Ibc) were found
to explode in early-type elliptical galaxies, while type II events were
found exclusively within star-forming galaxies (see e.g. \citealt{rea53} for an early example, confirmed by
many subsequent studies, e.g. \citealt{van05}, \citealt{hak14}). 
However, even further back in time \cite{baa34} made the observation that ``...super-novae 
occur not only in the blurred central parts of nebulae but also in
the spiral arms...". This was reference to the spatial
distribution of novae and supernovae within their host galaxies and hence environmental
information was already used through those early investigations 
to note differences in the properties 
between these two events.
\cite{joh63} concluded that all SN types were concentrated in the disks of galaxies.
Progenitor constraints from host galaxies were given further strength through research investigating the correlation
of SNe with spiral arms (thought to trace SF) within galaxies. 
\cite{maz76} demonstrated that SNe~II were much more concentrated 
within spiral arms than SNe~I (also see \citealt{hua87})\footnote{Note, while in these
studies there was probably a mix of SNe~Ia and SNe~Ibc, given their relative rates, most of the sample was
probably dominated by SNe~Ia, hence the significant difference in correlation.}, again implying
shorter lived progenitors for the former (\citealt{mcm96} later showed
that this difference indeed exists between pure samples of SNe~II and SNe~Ia). 
While spiral arms trace the general waves of SF within galaxies,
on a more local scale high mass on-going SF is traced by \hii\ regions. Hence, the next logical step 
was pursued and analyses were carried out investigating the association of different SN types with
these regions by \cite{van92,bar94,van96}. These authors estimated the different degrees of association
of SN types to \hii\ regions through measuring the distance to the nearest region from each explosion site.
At this epoch SNe~I had now been separated into SNe~Ia and SNe~Ibc, and the main conclusion from these studies
was that any significant difference between the correlation of SNe~II and SNe~Ibc (both arising from massive stars)
and \hii\ regions was not observed (which, with larger samples and a distinct analysis technique 
we show to be incorrect below).\\
\indent Another analysis technique used historically, is to investigate the radial distribution of different SN types, and compare these to our understanding of 
radial trends of e.g. metallicity in spiral galaxies. \cite{bar92} concluded there were no statistical differences
in the radial distributions of different SN types. However, later investigations by \cite{van97}, \cite{wan97} and
\cite{tsv04} showed that SNe~Ibc are generally more centrally concentrated within galaxies than SNe~II, while
SNe~Ia appeared to show a deficit of events within central regions. The central bias of SNe~Ibc
with respect to other SN types has historically been interpreted as a metallicity difference as the
central parts of spiral galaxies are generally host to higher metallicity stellar populations (due to 
metallicity gradients within galaxies,
see \citealt{hen99} for a review). However, as we will show below, this interpretation is complicated when
one delves deeper into specific host galaxy properties.\\
\indent In the above the reader has been brought up to date to the sparse `environment' studies initially undertaken
during the 20th century.
The rest of this review will go into detail on the `environment' studies of the last decade, where this
field has substantially matured. First, for the sake of the rest of the article, we will define 
what we mean by `environment' in the context of the current review.

\subsection{A definition of `environments'}
In this paper, we define SN environments studies as:\\
\begin{itemize}
\item Using observations when the light from the SN is not contaminating that of the parent
stellar population, i.e. either before the SN has exploded, or once it has faded below the 
luminosity of the environment light. This constraint is important in order to analyse 
the parent stellar population light uncontaminated by SN emission.
\item Use of observations of unresolved (into individual stars) stellar populations. This then rules
out analyses of resolved stellar populations in very nearby galaxies. While results from
such studies can be very revealing (and indeed will be discussed later), this constraint focusses the discussion on unresolved populations.
\item Analyses where one separates host galaxies into distinct regions. This distinguishes environmental
studies from global host work. While this factor is somewhat due to the researcher's own choice, it is also affected
by distance resolution effects. As one goes to larger distances it becomes more difficult to resolve  
individual galaxies into separate detector pixels.
\end{itemize}
To obtain progenitor constraints on SNe, one then assumes that stellar population properties are representative
of those of progenitor stars/stellar systems. Given the short lifetimes of CC SNe (at most several tens of Myrs, 
although note the possibility of longer lived CC SN progenitors produced in binary systems), this 
assumption is most likely valid. In the case of SNe~Ia, their longer delay times before explosion make
interpretations more problematic. There are also issues with chance alignments of SNe onto stellar populations
which are not associated with those of the progenitors. These issues and others will be further discussed
in \S\ 4.

\section{STATISTICAL TECHNIQUES}
Environmental analyses have been undertaken through both photometric and spectroscopic avenues.
Photometric analyses have focussed on the flux at the particular explosion sites through different
broad- or narrow-band filters, or multiple filters tracing environment colours (e.g. \citealt{kel12}). 
These observations
can be linked to particular properties of stellar populations, predominantly age, but also 
line of sight extinction, and metallicity (through stellar population modelling). 
Analyses of SN radial distributions, as normalised to stellar population distributions throughout
galaxies have also been undertaken.
Spectroscopic
analyses have generally concentrated on metallicity derivations through emission line ratio diagnostics,
however equivalent width measurements (see e.g. \citealt{kun13} and further discussion below), 
have also been used.\\
\indent The general methods of these studies will now be outlined, however the reader is encouraged
to delve deeper into individual references for a complete description of these techniques, their
possible biases and error estimations.

\subsection{Host galaxy pixel statistics}
SN environment studies started in earnest with the introduction of host galaxy `pixel statistics'. In 2006 
two papers independently developed very similar techniques with the aim of
analysing where within the distribution of pixel counts of particular galaxies 
the SN environment
falls. \cite{jam06} investigated the association of different SN types with host galaxy on-going SF, through 
narrow-band \ha\ observations, while \cite{fru06} analysed the association of LGRBs,
to host galaxy $g'$-band light as compared to CC SNe. The majority of the below discussion will
focus on the analysis using the \citeauthor{jam06} method, that will now be described in more detail. However,
the \citeauthor{fru06} method is very similar (and has also been used subsequently by e.g.  \citealt{kel08}).\\
\indent The description to provide pixel statistics (dubbed `NCR' values in \citealt{and08}) is
outlined below with respect to \ha\ imaging. However, this description is valid for analysis 
of host galaxy observations in any waveband (and indeed could be applied to spectral observations as
we will outline in \S\ 5).
To begin, continuum-subtracted \ha\ images are trimmed to remove sky regions outside the galaxy.
Then, pixel counts within the resulting image are ordered in terms of increasing count. From this
the cumulative distribution is formed.
This cumulative distribution is then normalised by dividing by the total count of all pixels.  
This then leaves negative values, followed by NCR values between 0 and 1. Negative NCR values, the
result of noise in the sky, 
are then set to 0. Thus this statistic is 
made such that values of 0 are equal to sky or zero flux, while a value of 1 means that the pixel
contains the highest flux of the whole galaxy. 
The final NCR value can thus be understood as the fraction of \ha\ flux arising from 
pixels with lower counts than the SN-containing pixel.
One can then proceed to measure NCR values for different 
SNe within different galaxies, and build statistical samples to be compared. If a distribution accurately 
follows the (in this example case) \ha\ emission, then one expects the distribution to be flat
with a mean NCR of 0.5. This would imply that in a population that accurately traces host galaxy
\ha\ emission, an equal number of SNe have NCR values in each bin of NCR (for more details
and discussion see \citealt{and12}).\\
\indent Using pixel statistics, one can use differences in the association of
SN types to tracers of different stellar populations to constrain progenitor properties. In the above example of \ha\ emission, 
this then gives differences in associations to very young stellar populations, and hence can be used to
probe relative age differences between progenitors, as will be shown below.\\
\indent The spatial resolutions probed by pixel statistics depend on the distance of the host
galaxies observed, together with the pixel scale of the instrument used and the image quality
of those observations. In the case of \ha\ NCR statistics, which will dominate
the discussion below, the median physical sizes probed are on the order of 300 pc (see \citealt{and12,and15} for 
more detailed estimations). 
Hence, in general these statistics do not probe individual \hii\ regions, but larger star-forming
regions within galaxies.

\subsection{Radial distributions}
A more generalised environmental parameter than that of the above pixel statistics, is the radial
position within hosts where SNe are found. Within galaxies the properties of 
stellar populations can change significantly with radial position. Metallicity gradients can exist
(see e.g. \citealt{hen99}), in that the central parts of galaxies are generally more metal rich
than outer regions. There also exist distinct populations in stellar bulges and bars within the 
centres of galaxies. Galaxy interactions and mergers can also significantly affect the SF processes within 
galaxies. Hence, one can analyse differences in the radial distributions of SNe and try to understand 
these in the context of the different stellar populations found at different galactocentric radii.\\
\indent The first studies to investigate the radial distributions of SNe generally normalised galactocentric
distances to e.g. $R_{25}$ (radius of galaxy to the 25th mag. isophote) or worked with absolute distances. One issue with this type of
normalisation is that it does not take into account the actual radial distributions of different
stellar populations. \cite{jam06} analysed the radial distributions of SNe with respect to 
those of the light as seen through distinct filters.
This analysis measures the ellipse with the same parameters as that of their host galaxy that just includes
each SN. One then measures the flux contained within this ellipse, and normalises to the flux contained within
an ellipse out to distances where the galaxy flux is consistent with sky values. One again obtains
values between 0 and 1, where a value of 0 would mean that a particular SN occurs at the central peak of 
emission, while a value of 1 means that the SN occurs out at distances where no significant galaxy flux is measured.
Building radial distributions of different SN types one can then analyse differences and try to tie 
these to differences in progenitor properties.
In the below discussion, we concentrate on SN radial distributions with respect to continuum $R$-band light, 
which is referred to as \textit{Fr}$_{\textit{R}}$. For a full discussion of these analysis
methods see \cite{jam06}, and \cite{and09}. There are several selection effects 
which may be at play in such an analysis, mostly due to missing SNe in the central parts 
of host galaxies due to the high surface brightness background. However, as has been 
shown in \cite{hab12}, there is no significant evidence that these issues will affect
one SN type preferentially over the other, especially in the case of differences
between SNe~II and SNe~Ibc. In \S\ 4.5 we discuss selection effects in general with respect 
to environment analyses.

\subsection{\hii\ region metallicities}
CC SN progenitor metallicity is predicted to play a significant role in the determination
of SN type, especially in the single star progenitor scenario (see predictions from e.g.
\citealt{heg03}). In extra-galactic studies, the simplest way to measure
stellar population metallicity is through observing the ratio of strong
emission lines produced through the ionisation of ISM material by massive stars. The metallicity
of choice is that of oxygen, as this is the most abundant metal found
in \hii\ regions
and shows strong emission lines in \hii\ region spectra. 
In addition, ISM oxygen abundances show
good correlation with stellar metallicities of massive stars (see e.g. \citealt{sim11}). 
While there now exist a plethora of oxygen abundance
diagnostics, there is significant disagreement between different
diagnostics, meaning that constraints on \hii\ region metallicities in absolute
terms are somewhat problematic. A detailed discussion of these issues, together
with suggestions of which diagnostic to use for specific cases is
given in \cite{kew08}. While oxygen abundance measurements between diagnostics can often
show large offsets, relative metallicities within a given diagnostic are thought to
be much more reliable. Hence, one can investigate \hii\ region metallicities between
different SN types, using a specific diagnostic.\\
\indent While the ISM metallicity presents the gas abundance at the epoch of observation, given the 
relatively short lifetimes of CC SNe (at most several 10s of Myrs) 
one can assume that the SN host \hii\ region metallicity is a reasonable tracer of 
progenitor metallicity. One issue is that it is common that at exact explosion sites
emission lines are relatively weak, meaning one has to observe the nearest emission line
region. Indeed, this is shown in \S\ 3.1, where particularly for SNe~II (but also for many SNe~Ib, and
a small number of SNe~Ic), CC SNe are found to explode within environments with relatively low
\ha\ flux. While it is not clear that this should bring any systematic bias to CC SN metallicity 
measurements, it is important to note the complications in direct progenitor metallicity
constraints from environments. Some studies have attempted to negate this problem
by only analysing metallicities which have been extracted at exact explosion sites.
However, if there is some progenitor effect which determines whether a SN is less or more
likely to explode within a bright \hii\ region (which we argue there is), then one 
may be simply replacing one bias with another. Hence, in our analysis below we include 
all `environment' metallicity measurements. 
We note that a relatively recent review of metallicity measurements of 
CC SNe and those SNe accompanying Long duration Gamma Ray Bursts (LGRBs), was also presented in \cite{mod11_2}.\\
\indent There have now been multiple studies of CC SN host \hii\ region metallicities 
\citep{mod08,and10,mod11,lel11,mod11_2,san12,kun13,kun13_2,sto13,tad13_2}. However, overall conclusions
on progenitor metallicities implied from these results are still not definitive (more
details from these studies will be discussed below).
In this review we present a meta-analysis,
combining all of the above measurements to try to further probe whether environment
metallicity differences between SN types indeed exist. 
The most common metallicity calibration used is that from \cite{pet04},
and hence we take those values for our analysis. Apart from being the commonly used
diagnostic (and hence published abundance values being readily available),
the PP04 metallicity scale also has the advantage that it is derived using emission lines
very close in wavelength: \ha\ together with \nii, and \hb\ together with \oiii.
This means that uncertainties in extinction corrections and/or relative
spectral flux calibration are unimportant. The PP04 scale
has two metallicity indicators: N2, using only \ha\ and \nii, and
O3N2, which additionally uses \hb\ and \oii\ line strengths. In a significant number
of cases publications only list N2 \textit{or} O3N2 values (the former because in many cases
one only detects \ha\ and \nii). As these two diagnostics are on the same scale (which
is tied to the ISM electron temperature), we use a combination of both values, preferring
O3N2, but using N2 when the former is not available.\\

\indent In the rest of this review we will concentrate on analysis and subsequent discussion of the results
obtained through the above methods. However, there are specific cases which are not covered 
by the above and these will
be introduced and discussed separately.

\section{SUPERNOVA ENVIRONMENTS}
Results will first
be summarised concentrating on the main CC SN types of SNe~II and SNe~Ibc, through pixel statistics, radial
distributions, and host \hii\ region spectral analyses. 
This is followed by some discussion of the environment studies which fall outside of these analysis
methods. Then SNe~Ia environments
will be discussed. We will make some comments on specific rarer SN sub-types, where
environmental analyses aid in simply deciding between a CC or thermonuclear origin.
Finally, we will present some results where environmental properties are linked to specific
SN properties, beyond classical spectral/light-curve classifications.

\subsection{Pixel statistics}
In Fig.\ 1 the \ha\ NCR pixel statistics cumulative distributions of: SNe~Ia, SNe~II, SNe~Ib and SNe~Ic 
are presented (taken from \citealt{and12}). These plots (which will be repeated in various forms below) 
can be understood in the following
way. As outlined above, if a population accurately traces the emission in which an analysis is presented --in the
case of Fig.\ 1 \ha\ emission-- then one expects that distribution to be flat with a mean of 0.5. On 
cumulative distribution plots this translates to being consistent with a straight diagonal line which
crosses the plot with intersections at 0,0 and 1,1. As one moves away to the upper left of this diagonal
(e.g. see the SNe~Ia in Fig.\ 1) then a population is displaying a lower degree of association to the emission.
If a population falls below this diagonal line to the right, then this equates to a distribution that is
biased towards the peaks of the emission, and does not simply follow the emission on a one-to-one basis.
Following this we see that the SNe~Ia show the lowest degree of association to the line emission, followed
by SNe~II, SNe~Ib and finally SNe~Ic showing the largest degree of association.
\ha\ emission arises from recombination of hydrogen ionised by the flux of nearby massive stars. Hence,
where one observes a concentration of \ha\ emission, an \hii\ region, one is in effect 
observing a concentration of young massive stars, where the dominant contributors to the
ionising flux are thought to be stars with initial masses $\ge$15-20\msun. 
Hence, these regions within galaxies are thought to have ages $<$15 Myrs (see e.g. \citealt{ken98}).
The distribution of \ha\ flux within galaxies therefore
traces on-going SF. Hence, if different SNe show differences in their association
to the line emission, as is shown in Fig.\ 1, then the most straightforward interpretation of this is that
we are observing differences in the mean progenitor ages, and therefore masses of SN classes.
Through this interpretation, we can then order SNe in terms of progenitor age and mass. The
results of Fig.\ 1 imply that, as expected, SNe~Ia have the longest progenitor lifetimes and
hence the lowest masses. SNe~II are next, and appear to have the lowest mass progenitors of
the CC population, consistent with constraints from direct progenitor detections. Then follows the
SNe~Ib, where pixel statistics suggest a slightly higher association to on-going SF than SNe~II, however we note that
this difference is marginal. 
The SNe~Ic meanwhile show an almost perfect one-to-one association 
to \ha\ emission. Hence, in this interpretation we have an increasing progenitor mass sequence 
of: SNe~Ia -$>$ SNe~II -$>$ SNe~Ib -$>$ SNe~Ic.
We note that there has been considerable criticism of this simple interpretation in the literature
(see e.g. \citealt{cro13,smi15}), and
we return to these issues later in \S\ 4. However, it is important to emphasise (irrespective of interpretations)
the key results from this analysis.\\
\begin{itemize}
\item Neither SNe~II nor SNe~Ib trace on-going SF as traced by \ha\ emission.\\
\item SNe~Ic accurately trace the spatial distribution of on-going SF within hosts.\\
\item SNe~Ib and SNe~Ic show a statistically significant difference in their association with on-going SF.\\
\end{itemize}
\indent At first glance, it may seem somewhat surprising that the SN~II population does not follow the spatial
distribution of \ha\ emission, given that the latter traces high mass SF, and that SNe~II are the explosions 
of massive stars. However, as outlined above, \ha\ emission traces SF on timescales of $<$15 Myrs, while the 
progenitor population of SNe~IIP has been estimated to have masses of $\sim$8--16\msun\ \citep{sma09}. Hence,
the latter have delay times before explosion longer than the lifetimes of typical \hii\ regions.
Near-UV emission traces SF on longer timescales of up to $\sim$100 Myrs (e.g. \citealt{gog09}). It has
been shown \citep{and12}, that indeed the SN~IIP population accurately traces the near-UV emission within
their host galaxies in terms of NCR pixel statistics, as is shown in Fig.\ 2. This is completely consistent
with these SNe arising from RSG progenitor stars at the low end of the CC SN mass range, as constrained
by direct progenitor detections.\\
\indent Similar work on pixel statistics has been published by other authors. \cite{kel08} presented
$g'$-band pixel statistics for SNe and compared these to long-duration Gama-Ray Bursts (these
authors used the fractional flux formalism of \citealt{fru06}, rather than the NCR method outlined
above). These distributions are shown in Fig.\ 3. While
\ha\ line emission within galaxies is dominated by contributions from on-going
massive SF, the $g'$-band light is a much coarser tracer of young populations. 
However,
it is interesting that similar trends to the above are seen in terms of the relative differences
between populations. SNe~II follow the $g'$-band light on an almost one--to--one basis, while the SNe~Ib
appear to be slightly more clustered on the brighter pixels. SNe~Ic meanwhile are much more concentrated 
on the brightest $g'$-band pixels (implied by the fact that the distribution falls to the bottom right of the 
diagonal). In absolute terms, this clustering of SNe~Ic (and to a lesser extent SNe~Ib) on the brighter 
$g'$-band regions perhaps implies that massive SF is clustered with the majority of it being found 
at the brightest continuum ($g'$-band) peaks. One of the main conclusions of \citeauthor{kel08}
was that SNe~Ic and LGRBs have similar mass progenitor stars due to their similar
associations to host galaxy young stellar populations as traced by $g'$-band light. These
same statistics were used by \cite{ras08} and compared to galaxy models to constrain CC SN progenitor
ages. It was concluded that the association of SNe~Ic to $g'$-band light implied a minimum 
progenitor mass for SNe~Ic of 25\msun.\\
\indent \cite{kan13} investigated specifically the local environments of CC SNe within infrared-bright galaxies.
Using NCR pixel statistics they presented similar distributions and relative differences in distributions
between SNe~II and SNe~Ic populations to the Anderson et al. results. Interestingly, these authors also presented pixel statistics
with respect to other wave-bands. In particular they showed that while SNe~II did not follow the $K$-band
light of their host galaxies, the SNe~Ic do. This is surprising given that historically $K$-band
light has been thought to trace the old stellar mass within galaxies. \citeauthor{kan13} suggested that
this was due to the fact that in starburst environments
the $K$-band light is dominated by red supergiants, and hence at least in their sample is a tracer of
SF. It is also interesting to note that with respect to the near-UV emission SNe~Ic cluster onto the brightest
pixels and show more than a one--to--one relation to the recent SF. This is also seen in Fig.\ 19 presented
below in \S\ 3.6.1, and may imply that the most intense SF happens in the brightest UV regions of galaxies.\\
\indent Finally, we note that \cite{cro13} and \cite{gal14} have also investigated the association of
SNe with SF regions, through pixel statistics and by measuring distances to the nearest \hii\ region.
Both authors find similar statistics to those that have been presented above: SNe~Ibc show a higher association
to on-going SF than SNe~II.\\

\begin{figure}
\begin{center}
\includegraphics[width=\columnwidth]{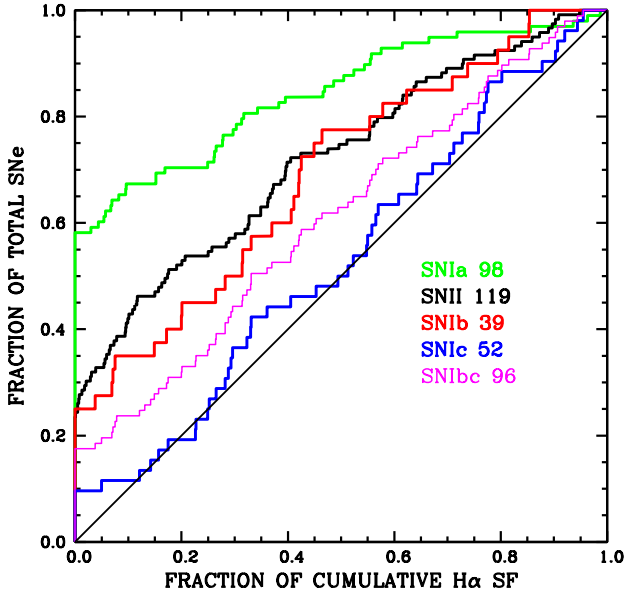}
\caption{Cumulative \ha\ NCR distributions of the main SN types: SNe~Ia, SNe~II, SNe~Ib and SNe~Ic (the combined
SN~Ibc population is also shown). The straight black diagonal line represents a hypothetical
distribution, infinite in size, which accurately traces (in this case) host galaxy on-going SF
as traced by \ha\ line emission. This plot is a reproduction of Fig.\ 2 from \cite{and12}, but here we
remove the SNe~IIb and SNe~IIn from the SN~II sample, to be consistent in our sample analysis throughout the 
review.}
\end{center}
\end{figure}

\begin{figure}
\begin{center}
\includegraphics[width=\columnwidth]{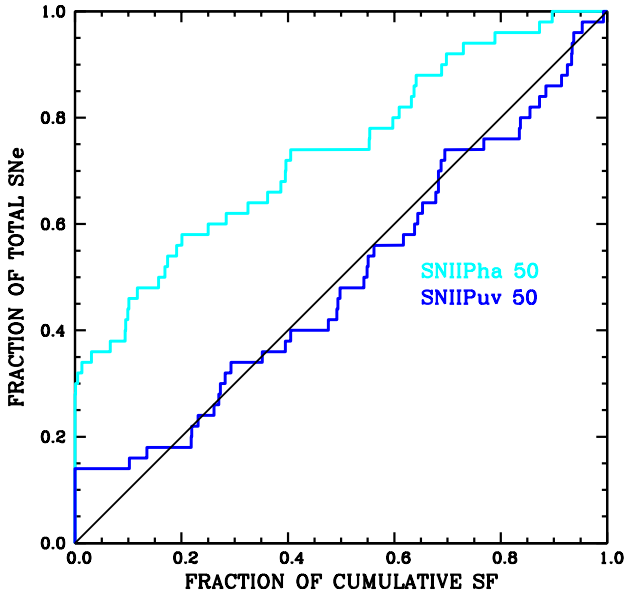}
\caption{Cumulative \ha\ and near-UV NCR distributions of SNe~IIP. 
Reproduction of Fig.\ 4 from \cite{and12}.}
\end{center}
\end{figure}

\begin{figure}
\begin{center}
\includegraphics[width=\columnwidth]{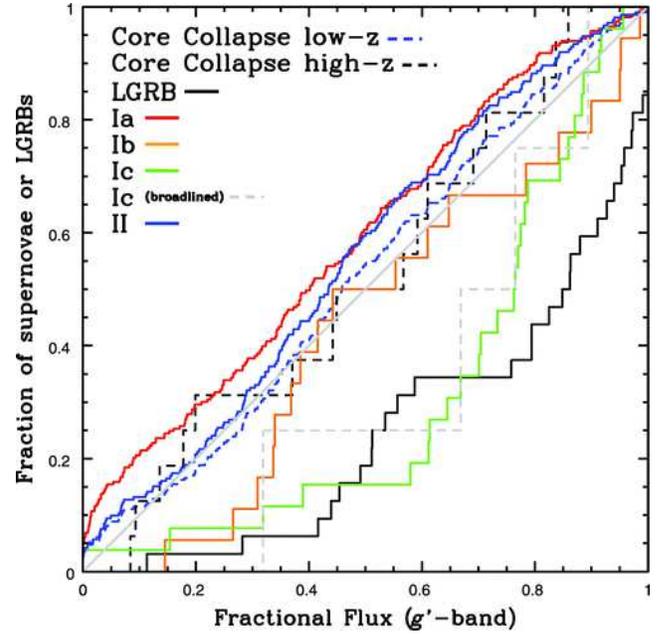}
\caption{Cumulative $g'$-band fractional flux distributions of SNe and long-duration
Gamma Ray Bursts. This figure is Fig.\ 2 from \cite{kel08}, and 
is reproduced with permission of the AAS. We thank Pat Kelly for consent to use
this figure.}
\end{center}
\end{figure}

\subsection{SN radial distributions}
An updated sample of radial distributions of CC SNe with respect to host
galaxy $R$-band light was published in Habergham et al. (2012, following \citealt{and09}, and
\citealt{hab10}, also see comparisons
to the radial distribution of SNe~Ia in \citealt{and15}).
That analysis concentrated 
on the radial distributions of the main CC SN types of
SNe~II, SNe~Ib and SNe~Ic. It has long been observed
that SNe~Ibc are more centrally concentrated within hosts than SNe~II (see e.g.
\citealt{van97} for an early example). Historically, this population difference
was interpreted as a progenitor metallicity effect (which was 
also the initial conclusion in \citealt{and09}), with a higher rate of SNe~Ibc
with respect to SNe~II in the higher metallicity central parts of galaxies.
In the single star progenitor scenario this is predicted to be the case,
as more metal rich stars have stronger radiatively driven winds (e.g. \citealt{pul96,kud00,mok07}), and 
therefore lose a greater proportion of their envelopes and hence explode as SNe~Ibc
rather than SNe~II. This conclusion has also been further supported by 
global host galaxy studies, where SNe~Ibc are found to occur in more massive galaxies 
than SNe~II (\citealt{pra03,boi09}, although see \citealt{arc10} for how this
is complicated when one separates SNe into their further sub-types), presumed to be
more metal rich, and higher metallicity galaxies through direct global measurements \citep{pri08b}.
However, \textit{when one separates host galaxies by their degree of
disturbance, metallicity as the dominant factor in explaining radial
distributions of CC SNe becomes untenable.}\\
\indent Separating CC SN host galaxies by their degree of disturbance was first achieved by \cite{hab10}, and a further analysis and discussion
was presented in \cite{hab12}. This classification separated host galaxies into
three groups: undisturbed hosts, disturbed hosts, and an extreme sample which is generally
dominated by major mergers (see examples of this last group in Fig.\ 8). 
Galaxy classification was achieved using a series of selection
criteria such as asymmetries, the appearance of a close companion, etc. The reader
is referred to \cite{hab12} for a full discussion.
Here, we add a further 67 CC SNe to the sample analysed in \cite{hab12}, including SNe
from the Carnegie Supernova Project (CSP), where we analyse host galaxy $r'$-band images 
originally obtained for light-curve host galaxy subtraction, and those SNe 
analysed in \cite{kan13} who employ the same method as \cite{and12}.
The final sample discussed here is of 347 CC SNe. This is split into:
226 SNe~II (where we include those SNe classified as `II', `IIP' and `IIL'), 50 SNe~Ib, 65 SNe~Ic and a total of 121 SNe~Ibc. 
We exclude from this analysis both SNe~IIb and SNe~IIn (see \citealt{hab14} for a discussion of the
radial distribution of SNe~IIn).\\
\indent For each SN in the above sample we calculate an \textit{Fr}$_{\textit{R}}$ which
gives a measure of where within the radial distribution of $R$-band light of their
host galaxies they exploded. The sample is then split by the degree of
disturbance, and we plot the resulting histograms and cumulative distributions in Figs.\ 4 through 6.
In Table 1 we give the mean \textit{Fr}$_{\textit{R}}$ values for each population in
each of the three galaxy samples, together with their overall values. Also presented
are the KS--test statistics which give `D' parameter indicating the amplitude of the 
difference between two populations and the probability `p' that the SNe~II and SNe~Ibc distributions
are drawn from the same parent population.\\

\begin{table*} \centering
\caption{CC SNe \textit{Fr}$_{\textit{R}}$ distributions and KS-test statistics}
\begin{tabular}[t]{ccccc}
\hline
\hline
SN distribution & Number of SNe & Mean \textit{Fr}$_{\textit{R}}$ & KS-test D& KS-test p\\
\hline
\hline
Undisturbed sample & &&&\\
SNe~II & 118&0.578 & &\\
SNe~Ibc& 51&0.483 & 0.224 & 0.048\\
\hline
Disturbed sample&&&\\
SNe~II & 85&0.489 & &\\
SNe~Ibc& 51&0.376 & 0.259 & 0.023\\
\hline
Extreme sample&&&\\
SNe~II & 23&0.507 & &\\
SNe~Ibc& 19&0.347 & 0.380 & 0.074\\
\hline	     
\end{tabular}
\setcounter{table}{0}
\caption{In column 1 the SN distribution type is listed, followed by the number 
of events analysed in column 2. Then the mean \textit{Fr}$_{\textit{R}}$ value
is listed for each distribution. For each set of two distributions the KS-test
`D' and probability value `p' are listed in columns 4 and 5 respectively.}
\end{table*}

\begin{figure}
\begin{center}
\includegraphics[width=\columnwidth]{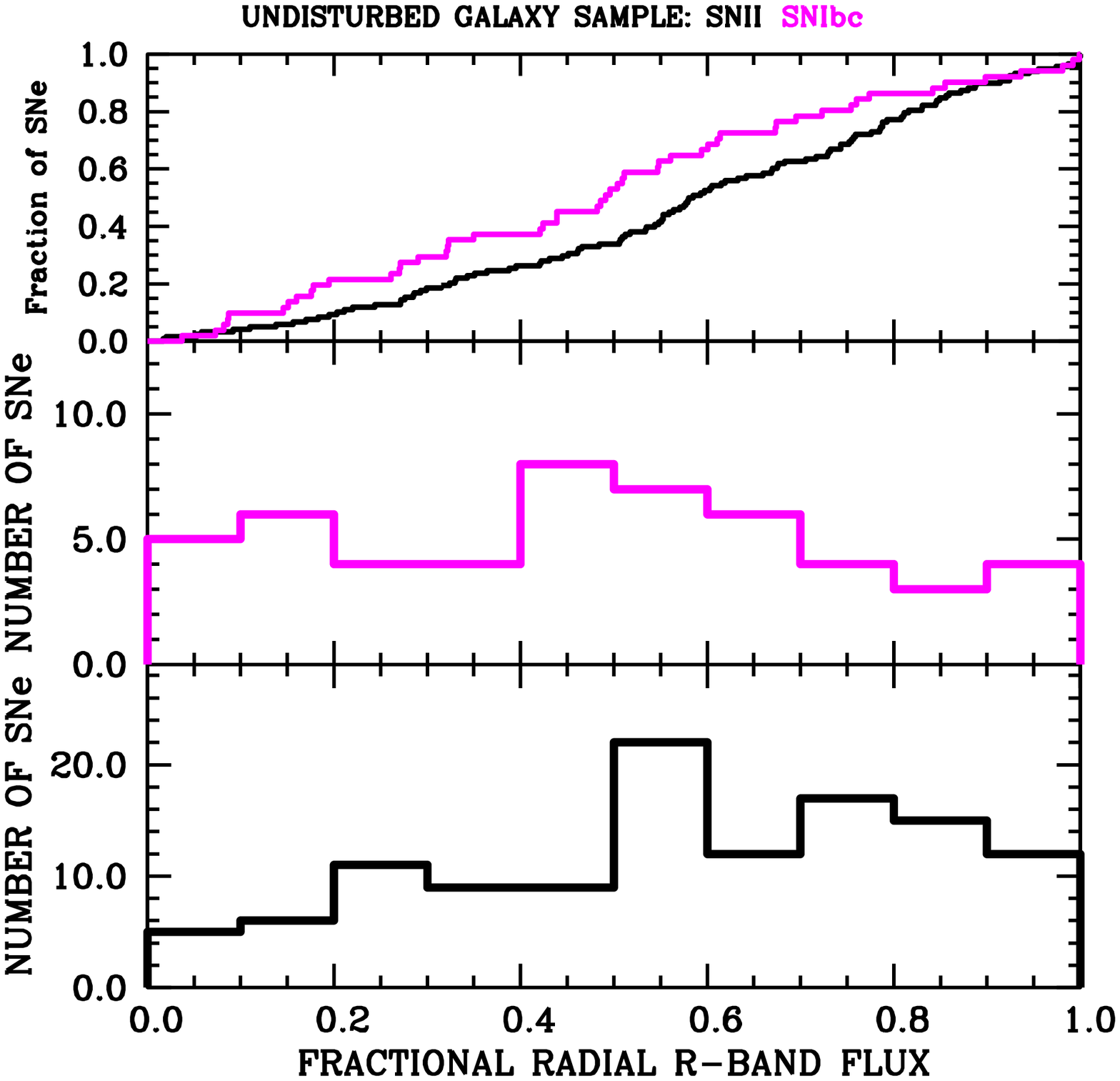}
\caption{Histograms of the \textit{Fr}$_{\textit{R}}$ distributions of CC SNe
in undisturbed galaxy systems. In the bottom panel the SN~II population is presented,
and in the middle panel the SNe~Ibc. In the top panel the cumulative distributions of
both populations are presented.}
\end{center}
\end{figure}

\begin{figure}
\begin{center}
\includegraphics[width=\columnwidth]{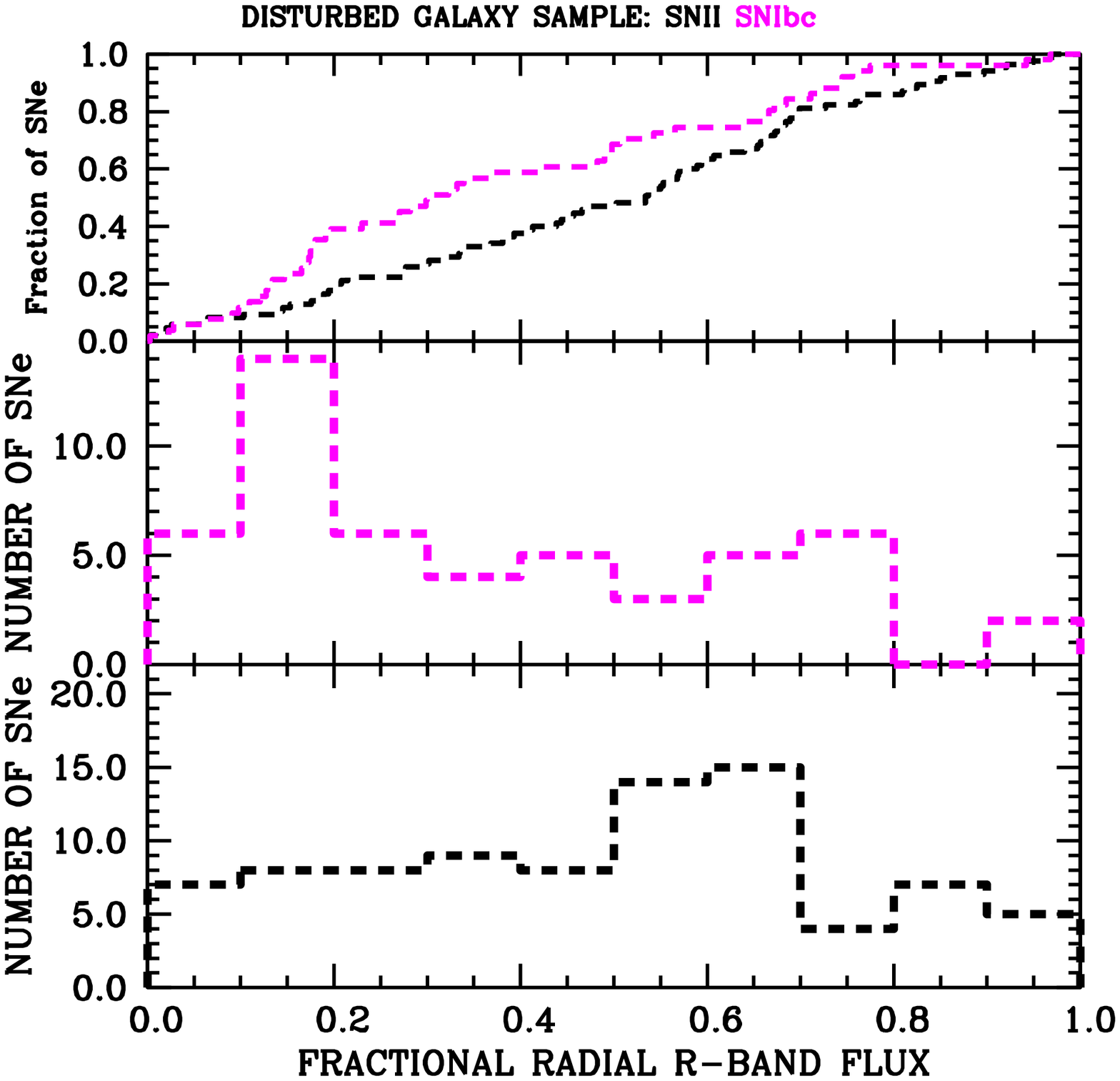}
\caption{Histograms of the \textit{Fr}$_{\textit{R}}$ distributions of CC SNe
in disturbed galaxy systems. In the bottom panel the SN~II population is presented,
and in the middle panel the SNe~Ibc. In the top panel the cumulative distributions of
both populations are presented.}
\end{center}
\end{figure}

\begin{figure}
\begin{center}
\includegraphics[width=\columnwidth]{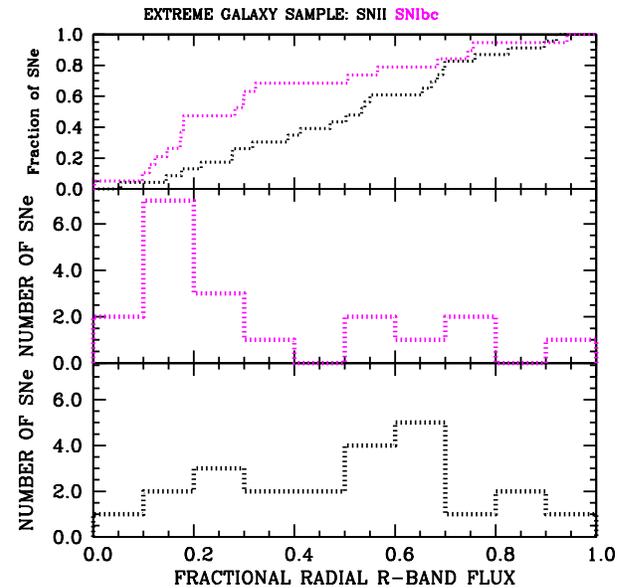}
\caption{Histograms of the \textit{Fr}$_{\textit{R}}$ distributions of CC SNe
in the most disturbed galaxy systems. In the bottom panel the SN~II population is presented,
and in the middle panel the SNe~Ibc. In the top panel the cumulative distributions of
both populations are presented.}
\end{center}
\end{figure}

\indent Analysing Figs.\ 4 through 6 and Table 1 we see the following trends. In all cases the SN~Ibc 
population is more centralised within host continuum $R$-band light than the SNe~II. 
However, \textit{this centralisation becomes more apparent} as one moves from the undisturbed through the
disturbed and finally in the extreme disturbed samples. This can be seen in the plots but also
in the KS-test `D' values in Table 1 which increase as one moves from undisturbed to disturbed host
galaxy samples. The figures show that this is mainly driven by an excess of SNe~Ibc in the centres of disturbed galaxies.
This is to say: as a galaxy sample becomes more disturbed SNe~Ibc become more centralised within 
their host galaxy light. It is also interesting to note the decrease in the ratio of SNe~II to
SNe~Ibc between the galaxy samples. In the undisturbed sample the ratio is 2.3, then 1.7 in the undisturbed
sample, and finally 1.2 in the extreme sample, i.e. the number of SNe~Ibc to SNe~II increases significantly 
as the degree of host galaxy sample disturbance increases. This is consistent with previous 
work \citep{hab10,hab12,hak14}.\\
\indent Throughout this review we show that there exist significant differences in the environments
of SNe~Ib and SNe~Ic. In Fig.\ 7 we present the \textit{Fr}$_{\textit{R}}$ distributions for the individual
SN~Ib and SN~Ic populations when the galaxy sample is split into undisturbed and disturbed galaxies
(an extreme sample is not discussed here due to the low number of events in that distribution). In Table
2 the mean values and KS--test statistics between these samples are listed. It is observed that here the 
difference between populations is more pronounced in the undisturbed samples (while above 
the difference between SNe~II and the overall SN~Ibc samples was more pronounced in the disturbed and
extreme samples). SNe~Ic are found
to occur more centrally than SN~Ibc in both undisturbed and disturbed galaxy samples, however the effect
is significantly larger in the former.

\begin{table*} \centering
\caption{SN~Ib and SN~Ic \textit{Fr}$_{\textit{R}}$ distributions and KS--test statistics}
\begin{tabular}[t]{ccccc}
\hline
\hline
SN distribution & Number of SNe & Mean \textit{Fr}$_{\textit{R}}$ & KS--test D& KS--test p\\
\hline
\hline
Undisturbed sample & &&&\\
SNe~Ib & 20&0.594 & &\\
SNe~Ic & 27&0.390 & 0.393 & 0.042\\
\hline
Disturbed sample&&&\\
SNe~Ib & 30&0.437 & &\\
SNe~Ic & 38&0.348 & 0.262 & 0.322\\
\hline	     
\end{tabular}
\setcounter{table}{1}
\caption{In column 1 the SN distribution type is listed, followed by the number 
of events analysed in column 2. Then the mean \textit{Fr}$_{\textit{R}}$ value
is listed for each distribution. For each set of two distributions the KS-test
`D' and probability value `p' are listed in columns 4 and 5 respectively.}
\end{table*}

\begin{figure}
\begin{center}
\includegraphics[width=\columnwidth]{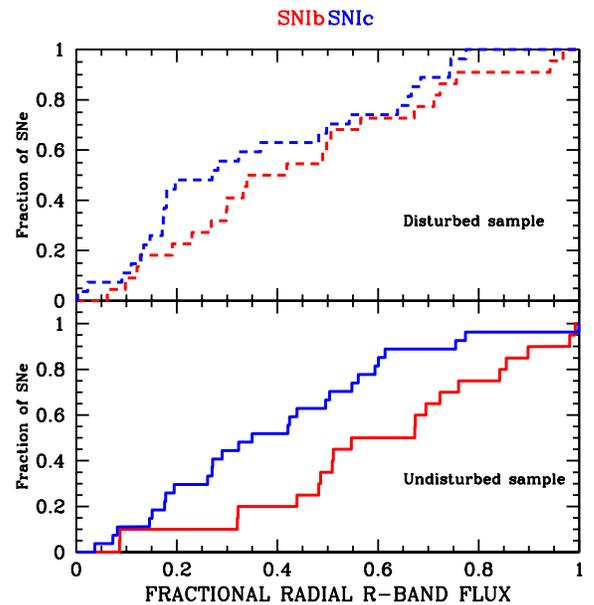}
\caption{Cumulative \textit{Fr}$_{\textit{R}}$ distributions for SNe~Ib
and SNe~Ic after host galaxies have been split into undisturbed and disturbed samples. (Note, here
we do not plot the separate extreme sample due to a lack of statistics, but this also forms
part of the disturbed sample.)}
\end{center}
\end{figure}

\indent While SNe~Ibc are more centrally concentrated than SNe~II
in disturbed systems, the explanation of this finding is not simple,
and it is not clear that metallicity effects are responsible. 
While
metallicity gradients are observed in most galaxies, they can become
much weaker or even non-existent in disturbed/merging galaxies (see e.g. \citealt{kew10} and most 
recently \citealt{san14}). During the
process of interaction pristine gas flows to the inner parts of the system, enriching
the central parts of galaxies with low abundance material (see e.g. \citealt{hib96,ram05,kew06}). In other
words, \textit{metallicity gradients are stronger in undisturbed systems, and hence
any progenitor metallicity effect between SNe~II and SNe~Ibc should be stronger 
in undisturbed systems than disturbed ones.} 
We also note that \cite{san14} showed that metallicity gradients can 
in fact invert in the very central regions,
which can further complicate the interpretation of SN radial distributions.\\
\indent When we separate the SNe~Ibc into Ib and Ic classes we see that the results are reversed. Now, the differences
between the radial positions are much larger in undisturbed systems (indeed
it is insignificant in disturbed systems). In this case the most likely cause is indeed
environment metallicity, with SNe~Ic preferring to explode in the higher metallicity central regions of galaxies
with significant metallicity gradients (consistent with the \hii\ region metallicity results presented below).
Before moving on we reiterate the main points from this radial analysis:\\
\begin{itemize}
\item Differences between the distributions of \textit{Fr}$_{\textit{R}}$ values of SNe~II and SNe~Ibc
\textit{increase} as one moves from undisturbed to disturbed host galaxy samples.
\item In disturbed/interacting systems SNe~Ibc are much more centrally concentrated within
galaxies than SNe~II.
\item There is a central excess of SNe~Ibc in the central regions of disturbed galaxies compared to host galaxy
continuum light.
\item Differences in the radial distributions of CC SNe \textit{cannot} be dominated by
progenitor metallicity effects.
\item However, in the \textit{undisturbed} sample SNe~Ic are seen to be more centrally concentrated than SNe~Ib, hinting
at a metallicity effect between these two sub-types.
\end{itemize}
\indent In \cite{hab10} and \cite{hab12} it was speculated on how these results can thus
be understood after discounting progenitor metallicity effects. Indeed, it was suggested that the centralisation
of SNe~Ibc in disturbed systems could indicate that in these dense violent SF processes induced
in the central regions of these galaxies, that the Initial Mass Function (IMF) is biased to the production of more massive
stars, and hence a higher proportion of SNe~Ibc to SNe~II is produced. It is also possible that
in the central regions of these galaxies there is a higher fraction of interacting binaries, and hence
a higher number of SN~Ibc explosions. Both of these possibilities were also considered 
in the radial analysis of \cite{hak09}, and are further discussed below.\\
\indent \cite{kan13} also analysed CC SN host galaxies using the same radial analysis as discussed above. 
These authors concentrated on SNe within infrared bright galaxies (many of which are starbursts), 
finding very similar results our analysis where we combine their sample into our large radial
analysis database. A further interesting result from \citeauthor{kan13} is that the SNe~II
in their sample were found to prefer to explode in the outer regions of the $K$-band light
of their hosts.\\

\begin{figure}
\begin{center}
\includegraphics[width=\columnwidth]{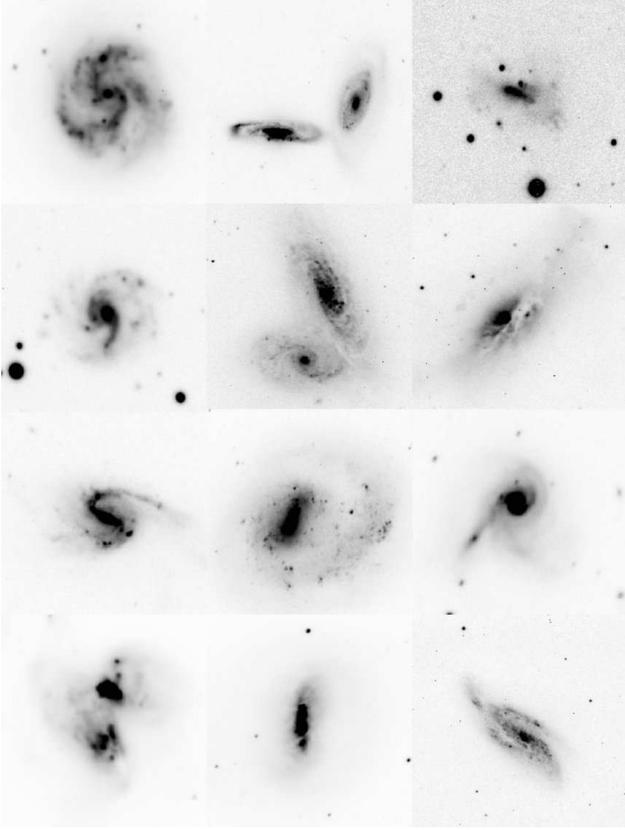}
\caption{Examples of the most disturbed galaxy systems where CC SNe
have exploded in the very central regions. This figure is Fig.\ 5 from
\cite{hab10}, and is reproduced with permission of the AAS.}
\end{center}
\end{figure}

\subsection{Host \hii\ region metallicities}
The first statistical study of SN environmental metallicities concentrated on
probing differences between broad-line SNe~Ic that had or had not accompanied
LGRBs, finding lower oxygen abundances for the latter \citep{mod08}. This suggests that
low metallicity is an important parameter in producing these extreme events.
In \cite{and10} a comparison was made between SNe~II and SNe~Ibc environment oxygen abundances.
It was concluded that any implied progenitor metallicity difference between these 
SN types was relatively small. That study also suggested that the difference between SNe~Ib
and SNe~Ic \hii\ region metallicities was small. However, this was contested by \cite{mod11},
who claimed a difference in that SNe~Ic arise from regions of higher oxygen abundance. 
These findings were interpreted as evidence for more complete envelope-stripping in
high metallicity progenitors, indicating a dominant role for line-driven winds. 
\cite{lel11} also presented an analysis of SNe~Ibc environment oxygen abundances finding
something in between those two previous studies.
A sample of SNe~Ibc formed from exclusively non-targeted SN searches was presented by \cite{san12}.
Those authors concluded that only a small difference between SNe~Ib and SNe~Ic environment oxygen abundance
existed,
and also analysed previous work in the literature to conclude that any difference between SNe~II
and SNe~Ibc \hii\ region oxygen abundances was small. A sample
of SNe~II from the un-targeted search program PTF (Palomar Transient Factory, \citealt{rau09}) 
was published by \cite{sto13}. This study looked in detail at the properties of
their sample, and found that it had a very similar
oxygen abundance distribution to SN~II samples drawn from galaxy targeted surveys (i.e. that in \citealt{and10}).
This is somewhat surprising, as it is generally assumed that a sample drawn from targeted searches will
be biased towards SNe in more massive, metal rich host galaxies, compared with those found within un-targeted rolling
searches.
Most recently \cite{kun13} and \cite{kun13_2} published SN~II and SN~Ibc samples of \hii\ region oxygen abundances,
where they focussed on very nearby events and analysed IFU spectroscopy, attempting to identify 
specific parent stellar clusters. Again, they found only marginal evidence for an implied
metallicity difference between the two SN samples, although
SNe~Ic were found to occur in regions of higher oxygen abundance than SNe~Ib.\\
\indent In light of the significant number or studies above,
we proceed to present a meta-analysis of all CC SN \hii\ region oxygen abundance measurements. We also
include unpublished values from the CALIFA survey, which will be presented in Galbany et al. (in preparation).
In our final sample we combine the unpublished values from CALIFA with
those from: \cite{mod08}, \cite{tho09},
\cite{and10}, \cite{mod11}, \cite{lel11}, \cite{and11}, \cite{pri12}, \cite{hab12}, \cite{van12_3}, \cite{san12}, \cite{sto13}, \cite{tom13}, 
\cite{kun13,kun13_2}, \cite{ins13}, and \cite{tad13}\footnote{We include all measurements
that can be considered of the `environment' of individual SNe. In practice this
means that we use values extracted at `the nearest \hii\ region', in addition to those
at exact explosion sites.}.
Our final sample is of 245 
CC SN host \hii\ region oxygen abundances.
Here we only choose to analyse SNe~II (120 SNe, where this includes SNe~IIP, SNe~IIL 
and those events classified as simply SNe~II
in the literature), SNe~Ib (56) and SN~Ic (59, with a total of 125 SNe~Ibc), to be consistent with the previous pixel statistics and
radial analyses presented above.
It is noted that here we include data from many different sources, where data reduction and 
spectral extraction may be different between samples. These samples also come from different
SN search sources, including targeted and un-targeted searches, which again brings issues such as
differences in relative ratios of SN types. However, our aim here simply
to bring all the relevant data together in one place, and use significant statistics
to investigate whether there indeed exist any statistical
differences in implied host \hii\ region metallicities.\\
\indent In Fig.\ 9 we present the cumulative oxygen abundance distributions of the main CC SN types.
All mean values, and KS-test D and p values
are listed in Table 3. It is observed that while essentially all distributions have statistically
very similar implied environment metallicities, the SN~Ic population has the highest mean value.
Statistically speaking \textit{we find no evidence for environment and hence progenitor metallicity
differences between SNe~II and the overall SN~Ibc population}. We note that there is a \textit{suggestion}
that SNe~Ic have higher metallicities than SNe~Ib, but this is not statistically significant.
In Fig.\ 10 we show the ratio of SNe~II to SNe~Ibc as a function of host \hii\ region oxygen abundance.
Here we simply split the sample into five equally-sized bins (in terms of numbers of SNe). Errors on ratios
are Poissonian. The ratio
is essentially flat with metallicity, with some suggestion that the SN~Ibc/SN~II ratio
decreases in the lower bin (which has a mean 12+log(O/H) of 8.29).
In Fig.\ 11 we further split the sample and show how the ratios: SN~Ib/II, SN~Ib/Ic, and
SN~Ic/II change as a function of host \hii\ region oxygen abundance.
The SN~Ib/Ic ratio appears to increase slightly with
decreasing metallicity. The SN~Ic/II ratio is flat with oxygen abundance, while the SN~Ib/II ratio appears
to increase slightly. Again we emphasise the low significance of all of these
trends. We list the important results from this CC SN \hii\ region metallicity analysis:
\begin{itemize}
\item Overall we find only small, statistically insignificant differences in the environment
oxygen abundances of SNe~II, SNe~Ib and SNe~Ic.
\item There is a suggestion that the ratio of SNe~II to SNe~Ibc increases in the lowest metallicity bin.
\item The SN~Ib to SN~Ic ratio appears to increase with decreasing oxygen abundance.
\end{itemize}
\indent These results suggest that progenitor metallicity is not a driving parameter in 
producing the  majority of CC SN diversity. This obviously does not rule out metallicity playing
a part in producing specific SN sub-types, or changing the properties of more specific 
transient features (other than simple type classifications). However, this does seem
to suggest that other parameters such as progenitor mass and/or the presence and influence of a binary
companion are much more dominant. Indeed, given 
that single stellar models (see \citealt{heg03,eld04,geo09,ibe13}) predict a strong dependence SN type production
with respect to progenitor metallicity, the above results (and those published in the literature 
previously, see \citealt{and10,san12} for combined SNe~II and SNe~Ibc analyses)
indicate that a significant fraction of CC SNe are the result of 
massive star binary interactions. We return to this conclusion in \S\ 4.\\

\begin{table*} \centering
\caption{CC SN \hii\ region oxygen abundances and KS--test statistics}
\begin{tabular}[t]{ccc}
\hline
\hline
SN distribution type& Number of SNe & Mean 12+log(O/H)\\
\hline
SNe~II & 120 & 8.57\\
SNe~Ibc& 125 & 8.59\\
SNe~Ib & 56  & 8.56\\
SNe~Ic & 59  & 8.61\\
\hline
SN populations & D & p\\
\hline	     
SNe~II--Ibc& 0.09 & $\gg$10\%\\
SNe~Ib--Ic & 0.21 & $\sim$15\%\\
SNe~Ib--II & 0.11 & $\gg$10\%\\
SNe~Ic--II & 0.13 & $\gg$10\%\\
\hline
\end{tabular}
\setcounter{table}{2}
\caption{In the top panel in column 1 the SN distribution type is listed, followed by the number 
of events analysed in column 2. Then the mean Z value
is listed for each distribution. In the bottom panel column 1 indicates
the two distributions being compared, and columns 2 and 3 list the KS-test
D and p values between those distributions.}
\end{table*}

\begin{figure}
\begin{center}
\includegraphics[width=\columnwidth]{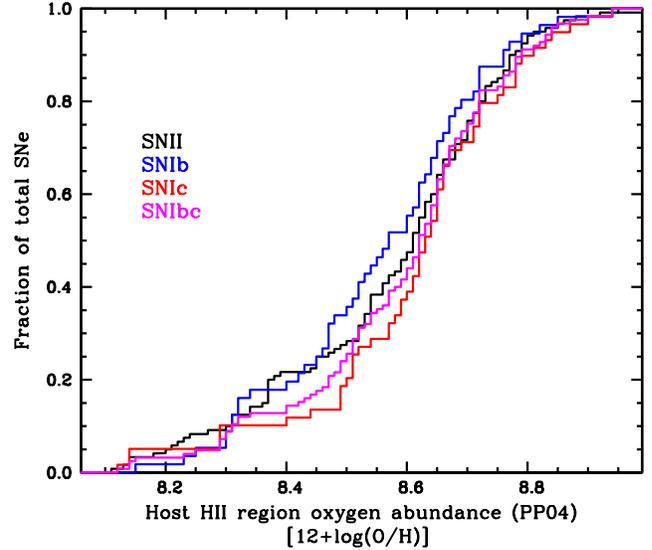}
\caption{Cumulative distributions of CC SN host \hii\ region oxygen abundance.
Here we show the SN~II, SN~Ib, SN~Ic and the combined SN~Ibc populations.}
\end{center}
\end{figure}

\begin{figure}
\begin{center}
\includegraphics[width=\columnwidth]{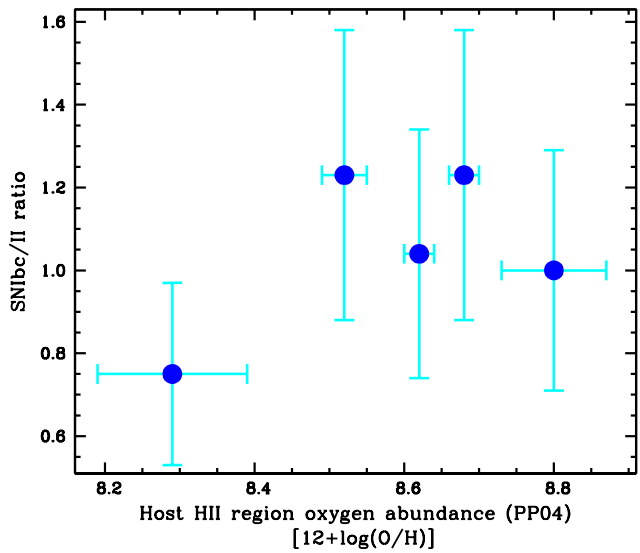}
\caption{The ratio of SN~Ibc to SN~II events as a function of host \hii\ region oxygen abundance.
The full sample of CC SNe are simply split into five bins of equal size (in terms
of number of measurements) of oxygen abundance.}
\end{center}
\end{figure}

\begin{figure}
\begin{center}
\includegraphics[width=\columnwidth]{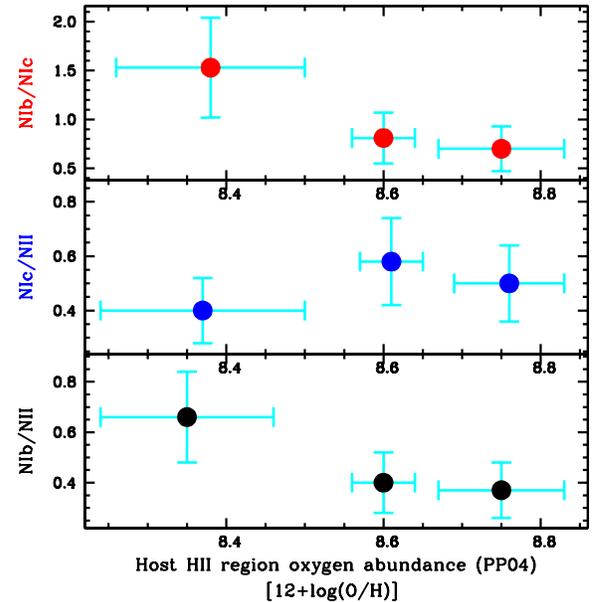}
\caption{\textit{Top:} the ratio of SNe~Ib to SNe~Ic as a function of host \hii\ region oxygen abundance;
\textit{Middle:} the ratio of SNe~Ic to SNe~II;
\textit{Bottom:} the ratio of SNe~Ib to SNe~II. In all three
histograms the samples are simply split into three equal sized bins of oxygen abundance.}
\end{center}
\end{figure}

\subsection{Further environment studies}
In this section we summarise and discuss CC SN environment studies which do not fall easily
into previous sections. Integral Field Unit (IFU) spectrographs allow
one to obtain both spatial and spectral information simultaneously. In \S\ 5
we discuss the future of environment studies where we argue that IFUs will play a dominant
role, following surveys such as CALIFA \citep{san12_3}, and using new facilities such as the recently
commissioned MUSE instrument at the VLT
\citep{bac10}.
\cite{kun13_2} and \cite{kun13} presented IFU environment studies of 11 SNe~II and
13 SNe~Ibc respectively. They used the power of IFU observations to first identify the nearest stellar cluster
to SN explosion positions, before extracting the spectrum at that position. Environment
oxygen abundances were then estimated using the same strong emission line diagnostics as outlined above
finding similar distributions for SNe~II and SNe~Ibc. These authors also measured \ha\ Equivalent Widths (EWs)
which they used to trace the age of the local stellar population, which can then 
be translated to progenitor masses using stellar evolution models.
In Fig.\ 12 we reproduce the mass--metallicity figure from \cite{kun13_2}, where measurements
were compared to both model predictions and SN~II direct progenitor detections.
It was concluded that a number of SNe~Ibc were associated with stellar clusters with
ages consistent with relatively low progenitor masses, where the 
implied progenitor scenario would be binary systems. In addition, in some cases
relatively high progenitor masses were derived for SNe~II. In Fig.\ 13
we reproduce the \ha\ EW distributions from \cite{kun13_2}, which show
differences between SNe~IIP, SNe~IIL, SNe~Ib and SNe~Ic. While the statistics
are limited, it appears that \ha\ EWs are generally similar between SNe~IIP and SNe~Ibc,
with some suggestion that those for SNe~Ic are slightly larger, possibly consistent
with the \ha\ pixel statistics presented above. The EWs of SNe~IIL appear to 
have a tendency for very large values, implying high mass progenitors. However, we caution that
these studies are currently limited by low statistics, where issues such as associating
a particular cluster to a parent progenitor population may become problematic. 
A significant number of new observations have been obtained for this study, and 
results from a larger sample promise to be enlightening.
The \citeauthor{kun13} studies concentrate on very nearby SNe where one is resolving individual
stellar clusters. \cite{gal14} used lower resolution IFU observations to also measure
\ha\ EWs at the explosion sites of SNe (and in addition analysed various measures of the association
of SNe to SF region, see below). In Fig.\ 14 \ha\ EW distributions for SNe~Ia, SNe~II and
SNe~Ibc are displayed. As expected, SN~Ia environments have the lowest EW distribution.
It is interesting that here the SNe~Ibc appear to explode in regions of slightly higher
EW than SNe~II, indicating younger ages, and therefore slightly higher masses for the former.
Again, larger samples and further analysis are needed to build on these studies.\\
\indent Another distinct environment analysis which has yet to be discussed, is
multi-colour photometric analysis. \cite{kel12} presented a large statistical
analysis of 519 CC SN environments deriving various parameters for the unresolved
stellar populations found at explosion sites. In Fig.\ 15
we present their environment $u'$-band surface brightness against $u'-z'$ colour plot.
SNe~Ic BL (broad-line) and SNe~IIb were observed to explode in significantly bluer
regions of their hosts than other SN types, while the $u'$-band surface brightness
of SN~Ic and SN~Ib environments was observed to be higher than those of SNe~II, again
somewhat consistent with the \ha\ pixel statistics above. 
These differences in environments suggest some differences in progenitor properties such
as age or mass (possibly related to $u'$-band surface brightness) and metallicity
(relating to environment colour).
Another interesting
result from \cite{kel12} was that no significant difference was found in the environment
properties of SNe~IIn and the `normal' SN~II population. This argues for similar progenitor
properties, a conclusion that will be further strengthened below.
\citeauthor{kel12} presented a number of other interesting environment and global host
galaxy measurements, and the reader is encouraged further explore that publication.\\

\begin{figure}
\begin{center}
\includegraphics[width=\columnwidth]{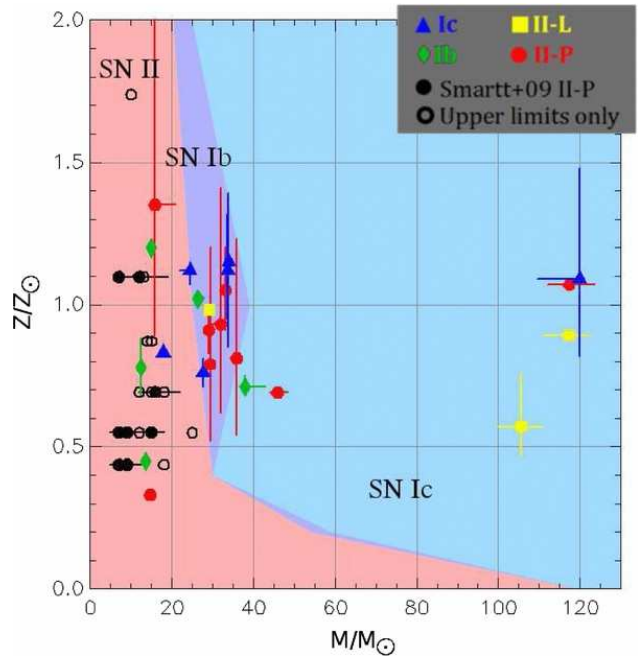}
\caption{Progenitor age against progenitor metallicity plot. SN environment ages and metallicities
are derived using IFU spectroscopy. These are compared to a model grid taken from \cite{geo09}, and
also SN~II direct progenitor detections from \cite{sma09}. This figure is Fig.\ 17 taken
from \cite{kun13_2} and is reproduced by permission of the AAS.}
\end{center}
\end{figure}

\begin{figure}
\begin{center}
\includegraphics[width=\columnwidth]{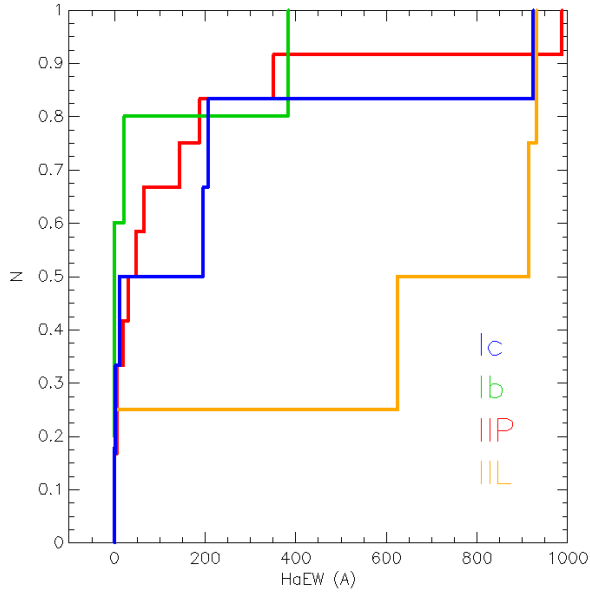}
\caption{\ha\ EWs of the stellar population observed within the parent stellar clusters of CC SNe, derived
from IFU spectroscopy. SN~IIP, IIL, Ib and Ic distributions are presented. A larger EW indicates
a younger age. This figure is an edited version of Fig.\ 18 taken
from \cite{kun13_2} and is reproduced by permission of the AAS.}
\end{center}
\end{figure}

\begin{figure}
\begin{center}
\includegraphics[width=\columnwidth]{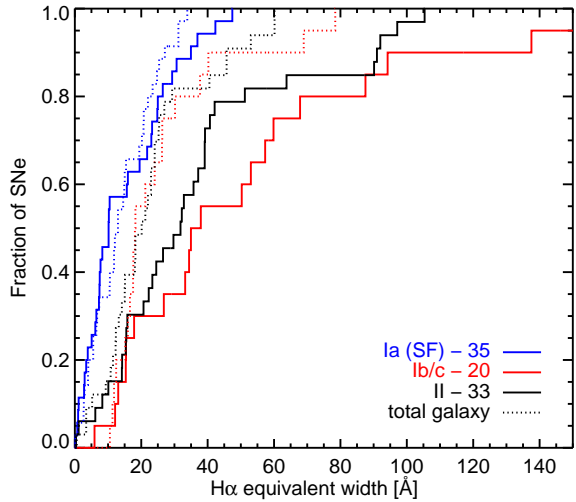}
\caption{\ha\ EWs of the stellar population observed at the explosion sites of SNe, derived
from IFU spectroscopy. SN~Ia, II, and Ibc distributions are presented. This figure is part of Fig.\ 13 taken
from \cite{gal14} and is reproduced by permission of A\&A.}
\end{center}
\end{figure}

\begin{figure}
\begin{center}
\includegraphics[width=\columnwidth]{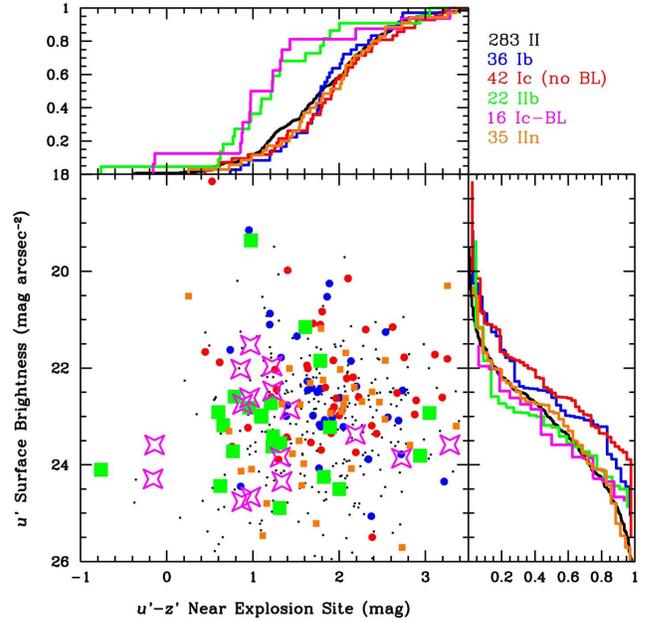}
\caption{$u'$-band surface brightness against $u'-z'$ colours 
of CC SN environments. The larger inset shows the two
parameters plotted against each other, while above and to the right are 
the cumulative distributions of each property. This figure is Fig.\ 2 taken
from \cite{kel12}, and is reproduced by permission of the AAS. We thank Pat Kelly
for consent to use this figure.}
\end{center}
\end{figure}

\subsection{SN~Ia environments}
This review concentrates on the environmental properties of CC SNe, as those
have been the most widely studied.
However, especially in the most recent years work has been published on the local
environments of SNe~Ia. Given the much longer expected delay times of SNe~Ia, 
interpreting environmental results becomes more difficult as progenitors have 
possibly travelled large distances before explosion. However, some intriguing
results have been published.\\
\indent \cite{ras09} used a modified version of the \cite{fru06} fractional flux method
to analyse SN~Ia environment pixel statistics, comparing observations to analytical galaxy models,
concluding that even the `prompt' component of the SN~Ia population has delay times
of several 100 Myrs. \cite{sta12} presented a first IFU study of SN host galaxies,
analysing the spatial and spectral information obtained for a sample of six SNe~Ia, and 
showing the capabilities of using such observations.
\cite{rig13} also used IFU observations of a large sample of SN~Ia host galaxies. These
authors concentrated on measuring the strength or presence of \ha\ emission at the explosion
sites of SNe~Ia. They found that SNe where emission was indeed detected 
have redder colours than those where no emission is detected (a similar result was
also found by \citealt{and15}, see below). 
\cite{wan13} speculated that two distinct populations of SN~Ia progenitors exist,
with those events having higher ejecta velocities being more centrally-concentrated
than their lower-velocity counterparts. However, this result has recently been questioned in an independent study by \cite{pan15}.
\cite{rig13} showed that SNe~Ia associated with local \ha\ emission are more homogeneous
and give a lower dispersion when used as distance indicators (also see \citealt{rig15}).
Subsequent work by \cite{kel15} identified a sub-set of SNe~Ia which exploded in environments of
high UV surface brightness and SF surface density, which gave significantly more accurate
distance measurements than SNe~Ia in all environments.
These results were predicted in the recent work by \cite{chi14_2}.\\
\indent In \cite{and15} the measurements
described in \S\ 2.1 and \S\ 2.2 are used to
study a sample of $\sim$100 literature SNe~Ia. 
Host galaxy NCR pixel statistics were obtained from near-UV through optical and near-IR imaging.
It was found that the SN~Ia population in star-forming galaxies most accurately 
follows the spatial distribution of host $B$-band light, as shown in Fig.\ 16.
This implies that the dominant progenitor population of SNe~Ia in star-forming
galaxies has neither very long delay times, i.e. they do not follow the near-IR $J$-
or $K$-band light, nor very short delay times, i.e. they do not follow the near-UV emission.
In addition it is shown in Fig.\ 17 that when the analysed SN~Ia sample is split by SN colour
at maximum, `redder' events are more likely to be associated with \ha\ emission
that their `blue' counterparts. This hints at host galaxy extinction driving 
the colour diversity of SNe~Ia. The reader is encouraged to read 
that paper in full for an overall view of SN~Ia environments.

\begin{figure}
\begin{center}
\includegraphics[width=\columnwidth]{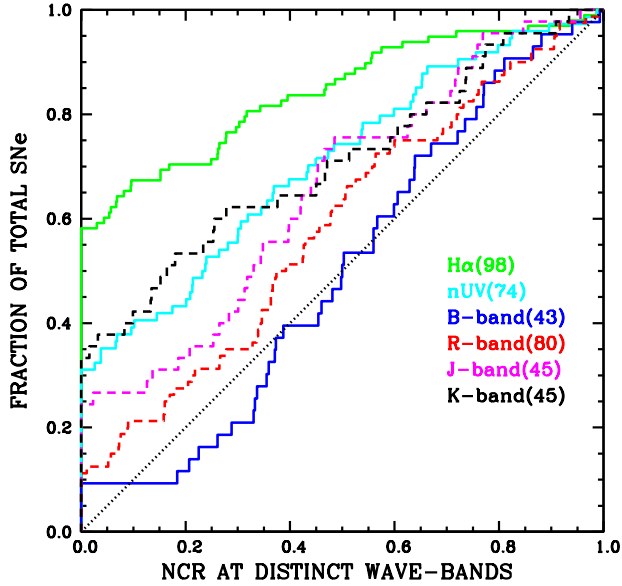}
\caption{Cumulative NCR pixel distributions for SNe~Ia at distinct wave-band observations: \ha,
near-UV, $B$-, $R$-, $J$- and $K$-band. This figure is taken from \cite{and15}.}
\end{center}
\end{figure}

\begin{figure}
\begin{center}
\includegraphics[width=\columnwidth]{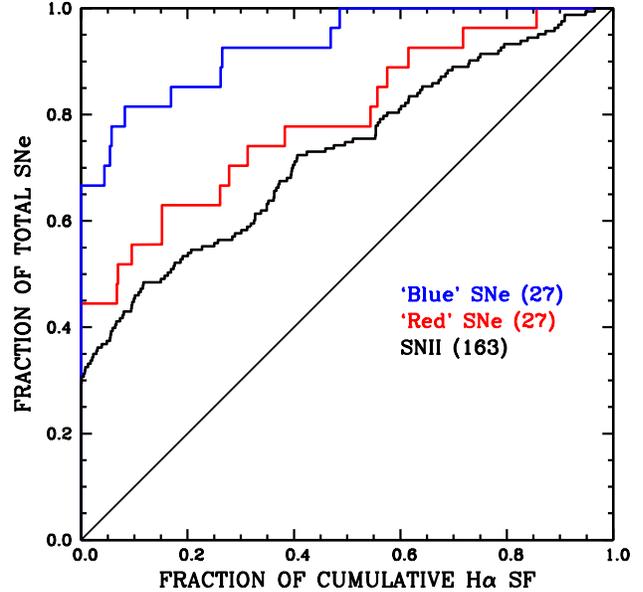}
\caption{Cumulative \ha\ NCR distributions for SNe~Ia when the sample is split by
SN colour at maximum light. The SN~II distribution is also shown for reference.
This figure is taken from \cite{and15}.}
\end{center}
\end{figure}

\subsection{The environments of SN sub-types}
Above we have focussed discussion on the environments of the main SN types.
In such studies one needs sufficient statistics 
to enable comparisons between different SN populations to be enlightening. However,
some studies have looked at (observed) rarer SN types and studied their environments 
with the aim of providing progenitor constraints. 

\subsubsection{Interacting transients}
SNe~IIn have historically been assumed to arise from extremely massive stars, given 
the large amounts of CSM material implied from their interaction-driven spectra. Indeed, it has been 
speculated (e.g. \citealt{smi08}) that their progenitors are Luminous Blue Variable stars (LBVs).
In addition, in the case of SN~2005gl the progenitor star was constrained to be a 
very massive and probable LBV star from detection on pre-explosion images \citep{gal09}.
In the case of the transient SN~2009ip (more details on the environment of this intriguing event below)
pre-eruption data appear to suggest a very luminous, very massive star.
If very massive star progenitors were the case for all SNe~IIn then one would expect 
their explosion sites to be similar to the environments of very massive stars or other
SNe expected to arise from such massive star populations.
In Fig.\ 18 we plot the cumulative \ha\ NCR distributions of SNe~IIn compared to those of other
SN types, as taken from \cite{hab14}. 
The surprising result is observed that SNe~IIn show the lowest degree of association
with host galaxy \ha\ emission of all CC SN types. This would appear to suggest that SNe~IIn have 
lower progenitor masses than previously assumed. A specific analysis of SN~IIn
environments is presented in \cite{hab14}. In Fig.\ 19 cumulative NCR pixel statistics are shown 
with respect to host galaxy near-UV emission. This shows that the SNe~IIn accurately trace the
recent SF within their host galaxies, in a very similar fashion to SNe~IIP. The similarity of the distributions
of SNe~IIn and SNe~IIP with respect to both \ha\ and near-UV emission, suggests that their
progenitor lifetimes, and hence masses are similar. In addition, in Figs.\ 18 and 19 
we also plot the SN~Ic population. The SN~Ic distributions are clearly offset from that of SNe~IIn, and 
if we assume that SNe~Ic arise from the highest progenitor masses of CC SNe, this is further
evidence that SN~IIn have relatively lower mass progenitors (we also note the differences
in their radial distributions seen in \citealt{hab14}). Another
interesting aspect of Fig.\ 19 is that SNe~Ic show a higher than one-to-one association  
to the near-UV emission. This is consistent with what was previously observed by \cite{kel08}
investigating SN pixel statistics with respect to host galaxy $g'$-band light
and again implies that the most massive SF is clustered to the brightest UV regions
within galaxies.\\
\indent There are a number of transients discovered where it is not clear whether they are the 
deaths of stars, or a large scale eruption which produces a bright (but significantly
dimmer than most SNe) transient. These transients have spectra very similar to SNe~IIn, and are named
SN `impostors' (see \citealt{van00,mau06}, and \citealt{koc12} for a recent discussion). The
environments of these events were also analysed and discussed in \cite{hab14}, and their pixel
statistics are shown in Figs.\ 18 and 19. Here we find that `impostors' show an even lower
degree of association to SF within galaxies than SNe~IIn. However, it is important to note that
the much lower luminosities of `impostors' compared to other SNe means that significant selection effects
could be at play in the discovery of these transients against bright \hii\ regions.\\
\indent SN~2009ip is a transient with an unclear nature. Indeed it is still debated whether
the star has finally terminally exploded or that the transient
is still being powered by eruptions and interactions \citep{smi10,pri13,pas13,mau13_2,fra13,mar14,smi14_2,gra14,mau14}.
One extremely interesting aspect of this transient that is often overlooked is the region
within its host galaxy where it exploded, i.e. its environment. As outlined in particular in
\cite{fra13}, SN~2009ip occurred in the very outer regions of its host galaxy, more then 4 kpc 
from the galaxy centre, with the vast majority of the host galaxy light 
being contained within the isophotal ellipse that passes through the SN position 
(indeed, \citealt{fra13} estimated an \textit{Fr}$_{\textit{R}}$
value of $\sim$1). This is shown in Fig.\ 20, reproduced from \cite{fra13}. 
While it has been shown in \cite{hab14} that SNe~IIn and `impostors' are generally found to more frequently
explode within the outer regions of hosts than e.g. SNe~IIP, SN~2009ip is still at the extreme 
of these distributions.
The explosion site of SN~2009ip
is somewhat surprising given that there are strong indications
that the progenitor was or is a
very massive star of $>$60\msun\ \citep{fol11}. However, there is no sign of on-going or recent
SF in the immediate vicinity of SN~2009ip, and neither is there any discernible spiral arm near to the 
explosion location \citep{fra13}. If the progenitor of SN~2009ip was such a massive star, as seems to be
the case, then its environment is puzzling. It may be that this transient is simply an example
of extreme isolated SF. It is also possible that the radial position of the explosion implies a low 
metallicity environment and hence progenitor. For now the environment of SN~2009ip remains a mystery, however
it will be intriguing whether such isolated regions within galaxies are typical 
of these types of events, once a statistical analysis is possible in the future.\\

\begin{figure}
\begin{center}
\includegraphics[width=\columnwidth]{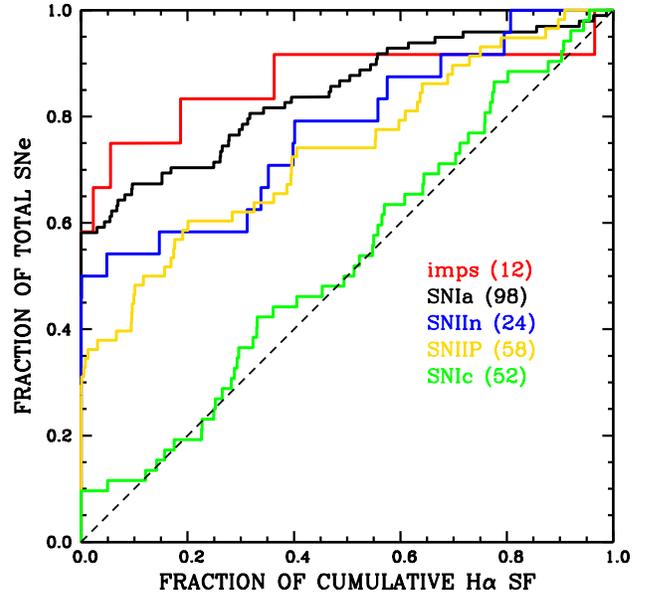}
\caption{Cumulative \ha\ NCR distributions of the interacting transients SNe~IIn and `impostors' as
compared to SNe~Ia, SNe~II, and SNe~Ic. This figure is Fig.\ 5 taken from \cite{hab14}.}
\end{center}
\end{figure}

\begin{figure}
\begin{center}
\includegraphics[width=\columnwidth]{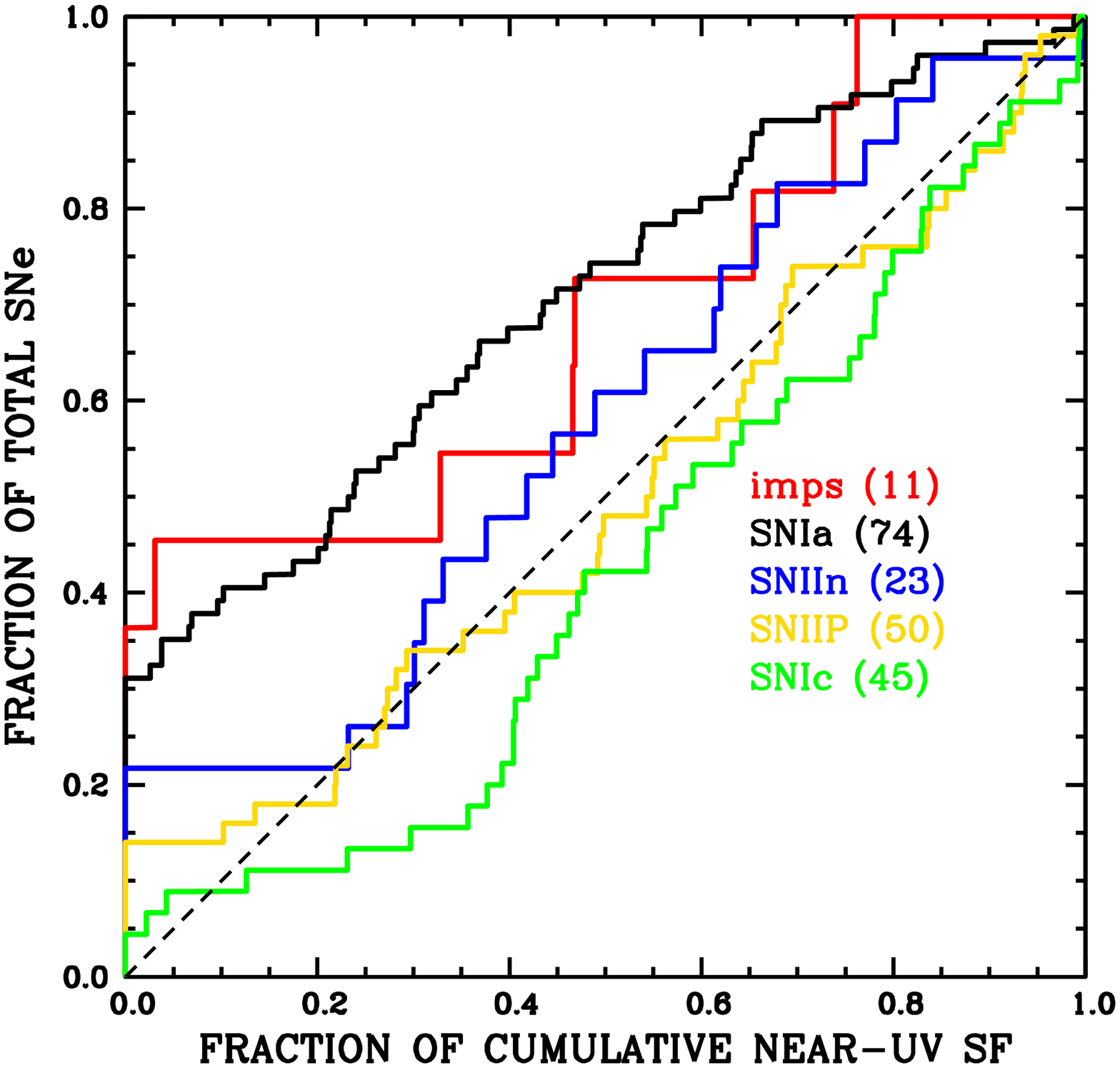}
\caption{Cumulative near-UV NCR distributions of the interacting transients SNe~IIn and `impostors' as
compared to SNe~Ia, SNe~II, and SNe~Ic. This figure is Fig.\ 6 taken from \cite{hab14}.}
\end{center}
\end{figure}

\begin{figure}
\begin{center}
\includegraphics[width=\columnwidth]{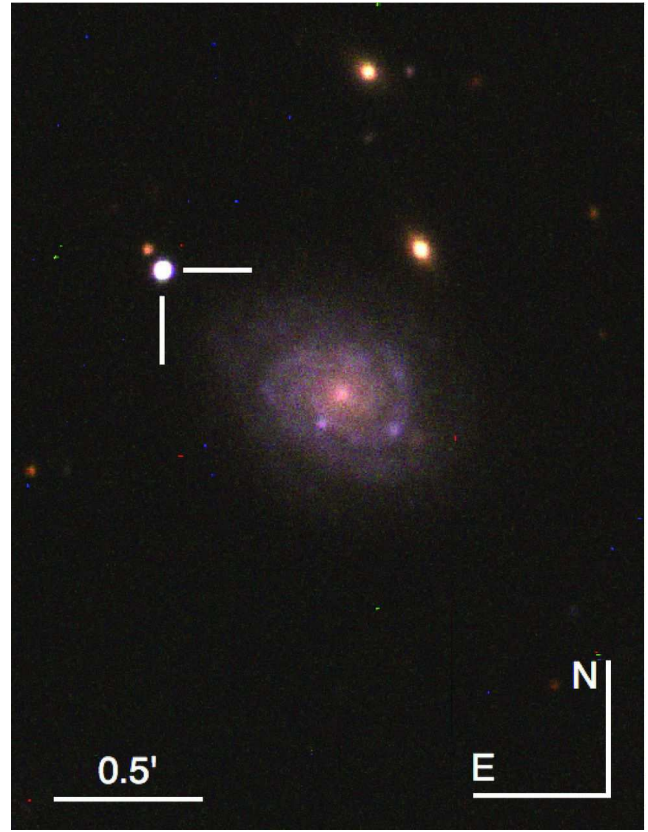}
\caption{The environment location of SN~2009ip within its host galaxy. This 
figure is a reproduction of Fig.\ 22 from \cite{fra13}.}
\end{center}
\end{figure}

\subsubsection{SNe~IIb}
As noted in the introduction, SNe~IIb are transitional events that first show signs of hydrogen (hence
the `II'), which then fades and the SNe appear similar to SNe~Ib.
Their relative rarity dictates that statistical studies of their environments 
lack significant numbers. However, some interesting trends have been observed.
In \cite{and12} it was found that SNe~IIb appeared to accurately trace host galaxy
SF, and there was a suggestion that the degree of association was higher than that
of SNe~Ib. In Fig.\ 15 it can be seen that SNe~IIb
have significantly bluer environments than most other CC SNe. In addition, there is
some suggestive evidence that SNe~IIb prefer to explode in lower metallicity
environments than the majority of other CC SNe \citep{mod11,san12}.
Hence, the environments of SNe~IIb do show differences to those of other CC SNe,
which may give clues as to their specific progenitor properties.
However, additional statistics are needed before concrete conclusions
can be made on progenitors.

\subsubsection{SNe~Ic BL}
A sub-set of SNe~Ic are known as SNe~Ic Broad-Line (BL), as their spectra show
significantly broader features, indicating significantly higher ejecta
velocities (and hence more energetic explosions). These events
are of particular interest because of their link to LGRBs \citep{woo06}. Studies of their
environments have shown that SNe~Ic BL usually explode in low-metallicity environments, 
in particular less metal rich than `normal' SNe~Ic \citep{mod11,san12}. However, those SNe~Ic BL
which accompany LGRBs appear to explode in even lower metallicity environments than SNe~Ic BL
without LGRBs (see \citealt{mod08} and \citealt{mod11}).
As shown in Fig.\ 15, \cite{kel12} demonstrated that SNe~Ic BL, like the SNe~IIb above, explode 
in blue environments, indicated by the $u'-z'$ colour at the explosion site.
\cite{kel12} speculated that for both SNe~IIb and SNe~Ic BL that the blue environments are linked
to the low metallicity of those same locations.

\subsubsection{87A-like events}
\cite{tad13_2} studied specifically the environments of SNe with similar properties to SN~1987A.
SN~1987A is the closest SN observed in the modern era of astronomy. SN~1987A was 
peculiar in that its light curve showed a slow extended rise (see e.g. \citealt{arn89}). There are now a handful
of similar SNe which have been discovered and studied \citep{pas05,kle11,pas12,tad12,arc12}.
SN~1987A itself exploded in the Large Magellanic Cloud at a distance of only 50 kpc. The wealth
of information from pre-explosion observations enabled detailed studies of its progenitor star
which was observed to be a blue supergiant star. Given the unusual nature of this SN, it is interesting to ask whether
these peculiarities are related to the specific environment within which it exploded.\\
\indent \cite{tad13_2} studied the environments of 13 87A--like SNe using very similar methods
to those outlined above. Fig.\ 21, from \cite{tad13_2}, shows the cumulative
distributions of the 87A-like population with respect to other CC SNe in terms of: host galaxy
absolute magnitudes; galacto-centric radial positions; and host \hii\ region metallicities. It 
appears that 87A-like SNe prefer environments 
with somewhat lower metallicities than other CC SNe,
as indicated directly by the measured oxygen abundances of their host \hii\ regions, and confirmed
by their fainter host galaxy luminosities and higher radial distances within these hosts.
\citeauthor{tad13_2} also investigated NCR values for their 87A-like sample, finding a similar distribution to
SNe~IIP.

\begin{figure*}
\includegraphics[width=17cm]{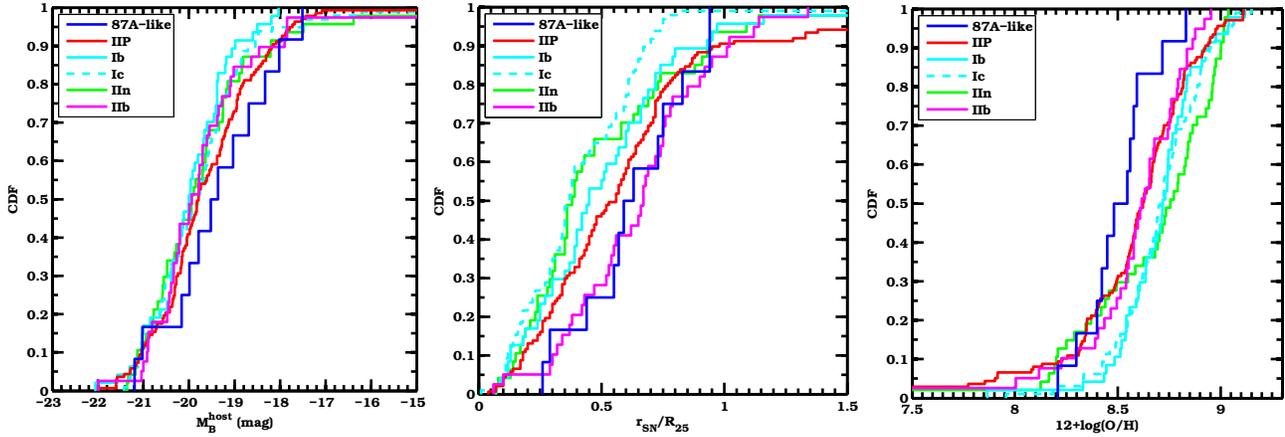}
\caption{Comparison of the environments of 87A-like events with other
CC SNe. \textit{Left:} cumulative host galaxy absolute magnitude distributions for
CC SNe. \textit{Middle:} cumulative $R_{25}$ normalised galacto-centric radial distributions of CC SNe. \textit{Right:}
cumulative host \hii\ region oxygen abundance distributions for CC SNe.
This figure is of Fig.\ 1 from \cite{tad13_2} and is reproduced by permission of A\&A.}
\end{figure*}

\subsubsection{SNe~Iax}
SNe~Iax is the term proposed by \cite{fol13}
for a set of SNe which have peculiar features
that strongly distinguish them from all other classes of SNe.
While they 
are spectroscopically similar to SNe~Ia, they have significantly fainter 
peak magnitudes and lower expansion velocities
at maximum light. Indeed, it is still not completely clear whether
these events are thermonuclear or CC in nature. Investigations into their environments
seem to imply relatively young progenitors. \cite{lym13}
showed that SNe~Iax have a similar degree of association to host galaxy on-going SF
as SNe~IIP, and used this to argue for progenitors with delay times
of less than 50 Myrs. It is interesting to note that
\cite{fol14} constrained the progenitor lifetime of the extreme SN~Iax SN~2008ha to be less
than 80 Myrs. Meanwhile \cite{mcc14} presented pre-explosion data of the SNe~Iax
SN~2012Z, and speculated that the progenitor was a white dwarf accreting matter from 
a helium star, however they could not rule out a massive star origin. 
While the interpretation of these detections remains unclear, both 
appear to point towards relatively short-lived progenitors for SNe~Iax, consistent
with their environmental properties.\\

\subsubsection{Ca-rich events}
SN~2005E was the first SN to be termed `Calcium-rich' (Ca-rich), by \cite{per10}.
A larger sample of these events have now been presented (see \citealt{per11,kas12,val14}).
Ca-rich events are usually spectroscopically classified as SNe~Ibc, but are significantly fainter
and present unusually strong calcium features. 
As above for the SNe~Iax, it has been debated whether these events arise from
massive star explosions (see e.g. \citealt{kaw10}), or thermonuclear events.
The strange locations of these
events was noted for SN~2005E \citep{per10} and SN~2005cz \citep{per11}, 
with both events occurring in non star-forming galaxies, and in the case of SN~2005E very far 
from their host galaxy. Indeed, in the compilation of \cite{kas12}, these remote locations
with large offsets from host galaxies were shown to be common.
\cite{yua13} compared the locations of Ca-rich events with results from 
cosmological simulations, concluding that their progenitors are likely 
to be of low metallicity and very old age. \cite{lym13} showed that the association of
Ca-rich events with host galaxy SF is similar to that of SNe~Ia, again favouring
significantly delayed progenitors. Most recently, \cite{lym14} presented
deep Hubble Space Telescope imaging 
of the explosion sites of Ca-rich events. The lack of any evidence of 
an underlying stellar population at the sites of these explosions led these
authors to conclude that Ca-rich SNe are in fact high velocity kicked systems, which have
moved with significant velocities from their birth sites, hence explaining
their environment properties.\\

So far we have not discussed differences in the environments of SNe~IIP and
SNe~IIL. Results have indeed been published on the environments of SNe~IIL (see e.g.
\citealt{and12,kun13_2}). However, type IIP/IIL classifications in 
the literature are extremely subjective. In the next section we present discussion on relating
environment properties to exact light-curve and spectral measurements of SNe~II.\\
\indent Finally, we note that we have not discussed the environments of Super-Luminous SNe (SLSNe). 
While such studies are certainly interesting, 
most of the published work thus far has concentrated on global host properties
(see e.g. \citealt{nei11,lun14,lel14,che14}), although a few very recent studies have investigated their immediate environments
\citep{lun14_2,tho14}. It will be interesting in the future to apply the same 
methods as used above to study these mysterious explosions.

\subsection{Correlating specific transient features with environment properties}
To date, environment and host galaxy studies have generally concentrated
on investigating differences in the environmental properties of SNe when events are simply split
by spectroscopic type. These typings are often somewhat subjective, and in the case of SNe~IIP and 
SNe~IIL, spectroscopic typing without any further light-curve information 
used can simply be wrong. This is an issue with SN catalogues which are often not updated
after initial classifications are given in circulars. Indeed, it has been shown in \cite{gut14} that
while correlations between SN~II spectral parameters and light-curve morphologies do 
exist, there is significant dispersion, which can lead to ambiguous spectroscopic classifications.
In the case of SNe~IIb and SNe~Ib, the epoch at which the classification spectrum is taken
can then lead to an event being typed into one class or the other (see e.g. \citealt{mil13}
for discussion on this issue). Given these ambiguities it makes sense for future
statistical studies to use more specific SN features to investigate environment
trends. While historically this type of analysis was difficult due to the lack of 
large SN follow-up databases, recent years have seen a number of publications of 
statistically significant samples of CC SNe (SN~Ia large samples have been relatively more widespread,
as is seen in the analysis in \S\ 3.5). SN~II photometry, spectra and measured parameters for large
samples have been
recently published in: \cite{arc12,and14a,and14b,gut14,far14a,san15_2,far14b}. SNe~Ibc and SNe~IIb
samples have been published in \cite{dro11,bia14,mod14}. This provides an opportunity to correlate
environment properties to more specific features such as light-curve morphologies,
the presence and form of specific spectral features and their evolution in time,
and parameters that have a direct link to explosion and progenitor properties such as synthesised nickel
mass.\\
\indent In this section we analyse the environmental properties of the SN~II sample presented
and analysed in \cite{and14a}. That work presented a sample of more than 100 well
sampled $V$-band light-curves from which key parameters were defined then measured.
A full analysis of all parameters and all possible environment measurements will be left for 
future publications. Here we present a few interesting figures using the analysis methods
outlined
in previous sections. In \cite{and14a} a parameter `$s_2$' was defined to give the decline
rate of SNe~II during their `plateau' phases (which are not necessarily flat, see that publication
for in-depth discussion). Using such a parameter allows one to move away from 
simple `IIP' or `IIL' classifications.
In Fig.\ 22 we present near-UV NCR distributions for SNe~II when the sample is split into
low and high $s_2$ populations (slow and fast decliners). One can see that the faster
decliners show a higher degree of association to the recent SF within galaxies.\\
\indent Host galaxy SN extinction is a key parameter in obtaining intrinsic SN brightnesses,
however especially in the case of SNe~II, a clear method for estimating
this parameter is not forthcoming (see e.g. discussion in \citealt{far14a}).
One measurement which is often used to define the presence of
host galaxy $A_v$ and its amplitude is
the EW of narrow sodium lines within spectra. However,
the accuracy of this measurement is debated (see discussion in \citealt{phi13} 
as applied to SNe~Ia). It can be thus enlightening to see how sodium absorption within SNe~II
spectra changes as a function of environment. In Fig.\ 23 \textit{Fr}$_{\textit{R}}$
distributions are plotted for SNe~II which do/do not show the presence 
of narrow sodium absorption within their spectra. One can see that those events 
with detections occur at smaller galacto-centric radii than those without.
This is to be expected as within the more central regions within galaxies 
a higher degree of ISM extinction is expected.\\
\indent Finally, in Fig.\ 24 we show SN~II \hii\ region metallicities, when the sample
is split by the \cite{and14a} parameter OPTd which is a measure of the optically thick
phase duration of the light-curve and measures the time 
between explosion and the end of the `plateau' stage. 
It is generally assumed that the length of this light-curve phase is related
to the mass of the hydrogen envelope at the epoch of explosion.
While the statistics
are small, it is observed that SNe with longer OPTd phases appear 
to explode in higher oxygen abundance regions than their shorter OPTd counterparts.\\

\begin{figure}
\includegraphics[width=\columnwidth]{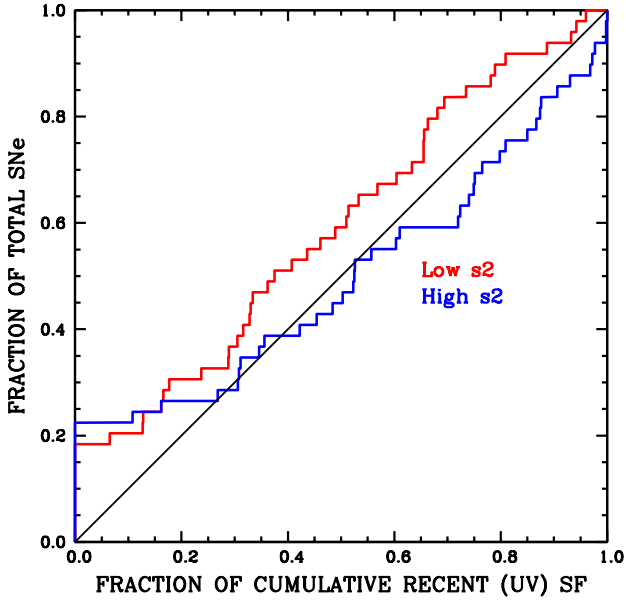}
\caption{Near-UV NCR distributions for SNe~II when these SNe
are split by their light-curve decline rate parameter: $s_2$. The latter
values are taken from \cite{and14a}.}
\end{figure}

\begin{figure}
\includegraphics[width=\columnwidth]{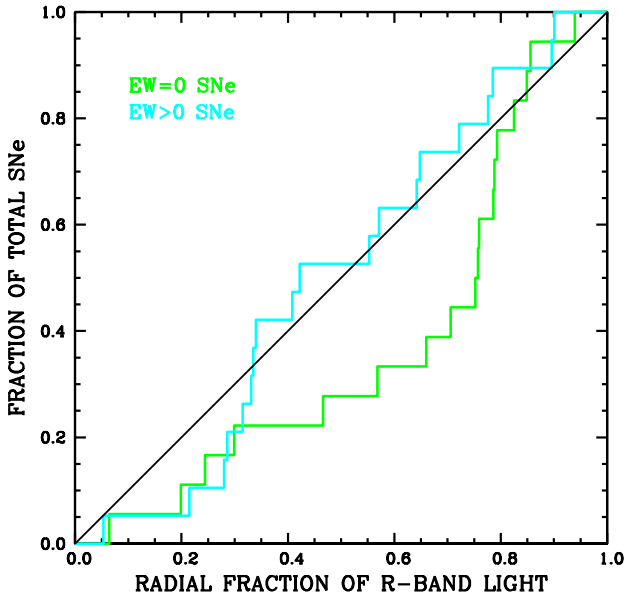}
\caption{\textit{Fr}$_{\textit{R}}$ distributions for SNe~II with and 
without narrow sodium absorption detections within their spectra.}
\end{figure}

\begin{figure}
\includegraphics[width=\columnwidth]{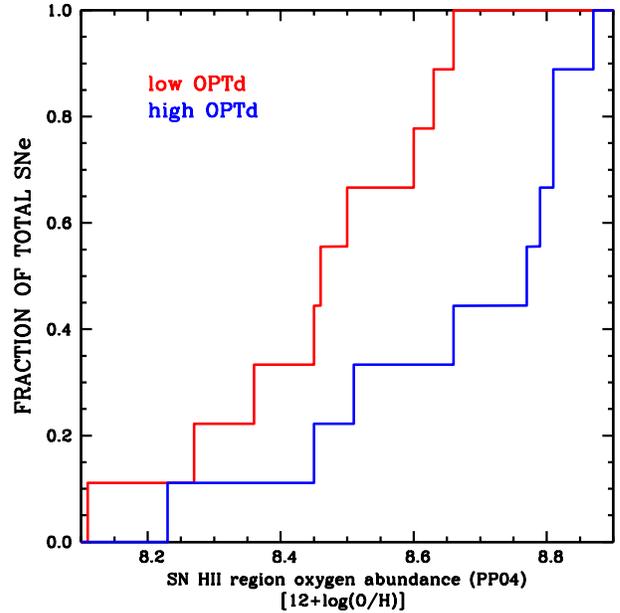}
\caption{\hii\ region metallicities for SNe~II when split into equal
samples of OPTd, the optically thick phase durations of their
light-curves. The latter
values are taken from \cite{and14a}.}
\end{figure}

\section{DISCUSSION}
In the above we have reviewed current results of SN environment studies, and shown
that significant differences exist in the properties of the explosion 
sites of different SNe.
These differences need 
to be explained by any proposed progenitor--transient mapping scenario.
Environmental results provide constraints on SN population progenitor properties
such as lifetime, mass and metallicity, where these also need to be tied to
the high probability that a significant fraction of CC SNe arise from binary systems.
In this section we outline how environmental
results can be used to infer progenitor properties, and also discuss issues with
the interpretation of the above results. Before moving to that discussion we note an
important point with regards to environmental studies. These investigations are statistical
in nature, in the sense that any single measurement, be it association to an \hii\ region, an emission
line metallicity measurement, an environment colour etc, does not necessarily imply 
a strong progenitor constraint on each individual SN. This is the case because of limiting factors such
as superposition of foreground or background stellar populations onto that of 
the specific SN, line of sight extinction which may mask specific environment properties, 
and progenitor velocities (from
e.g. kicks due to previous SN explosions within a binary system). 
However, statistical studies allow one to overcome these limitations as one can look at how 
different SN populations compare in terms of overall environmental properties. Hence, if a 
SN population has a mean e.g. metallicity higher than another, then this constrains
the overall population of the former to have higher progenitor metallicities than the latter. This
may seem like a pedantic point, but it is important to keep in mind when discussing statistical
constraints together with individual transient properties.\\

\subsection{Environmental constraints on SN progenitor mass}
We have shown above that SN types have an increasing degree of association to
host galaxy SF, starting from SNe~Ia, through SNe~II, SNe~Ib and finally SNe~Ic 
(an observational result which has also been observed with various degrees
of statistical significance by \citealt{kel08,cro13,kan13} and \citealt{gal14}). 
A major assumption in tying this to progenitor constraints is that
a larger degree of association of any given SN type to especially the youngest on-going
SF traced by \ha\ emission, implies shorter stellar lifetimes and hence more massive progenitors.\\
\indent The vast majority of stars form in clusters (\citealt{lad03}, although see
\citealt{bre10}), while
massive stars are generally born in clustered environments (\citealt{por10}, but see
\citealt{wri14}).
Massive young stars provide the ionising flux to create \hii\ regions. As time passes the most massive stars within 
a cluster explode as the first set of SNe. When there are no surviving stars of sufficient mass then
an \hii\ region will cease to exist. In addition as time passes clusters will tend to dissipate, and hence 
longer lived progenitors will be less likely to be associated with their parent cluster.
The passage of time also enables progenitors to move away from their parent cluster 
if they have some peculiar velocity with respect to the cluster. As time passes
gas will also be removed from a cluster due to the winds of the most massive stars, together
with the first set of SN explosions.
A full discussion of these 
processes (and others) can be found in \cite{and12}. Following this, the simplest interpretation of Fig.\ 1 is that
SNe~Ic have shorter lifetimes than SNe~Ib, which in turn have shorter (but quite similar) lifetimes
as the progenitors of SNe~II. Hence, SNe~Ic have higher progenitor masses than SNe~Ib, which have higher masses
than SNe~II.\\
\indent It is important to note here, our argument for higher
mass progenitors for SN~Ic as compared to SN~Ib and SN~II does not necessarily constrain the former
progenitors to be single stars. It simply states than on average SN~Ic have shorter lifetimes and
more massive progenitors than SN~Ib and SN~II, irrespective of the progenitor scenario. 
Indeed, this interpretation gains weight from the age constraints published by \cite{gal14}, and shown in
Fig.\ 14 (we also note that the higher resolution investigation of \citealt{kun13}
gave qualitatively similar results, however conclusions are less clear due 
to small number statistics).
In the case of SNe~Ib, the fact that they show a very similar degree of association to on-going
SF as SNe~II, indicates that the SN~Ib progenitor population is dominated by binary systems. This is because
if SNe~II and SNe~Ib arise from similar mass progenitors (implied from their NCR 
distributions), then single star mass-loss processes
are insufficient to remove the entire hydrogen envelope. Mass transfer between
binary companions can explain this result, and the importance of binary systems 
in understanding CC SN progenitors is discussed in detail below.\\
\indent \cite{cro13} analysed very nearby CC SN
positions with respect to \hii\ regions, and in particular discussed the ages of both individual
isolated regions, and large giant nebulae. This author found that SNe~Ibc were more associated
to star-forming regions than SNe~II, consistent with what has been outlined above. Indeed, \citeauthor{cro13}
also concluded that this implied higher mass progenitors for SNe~Ibc than SNe~II (we re-iterate that this does
not necessarily differentiate between single and binary progenitor scenarios). An important difference between that work and the NCR
pixel statistics above is the lack of statistics which did not allow \citeauthor{cro13} to separate
SNe~Ib from SNe~Ic. When one obtains sufficient statistics to enable this separation, there is a clear
difference between SNe~Ib and SNe~Ic in terms
of their preference for exploding near to \hii\ regions, with the latter showing
a much higher association. Indeed, this is a key observational result which needs to be explained
through progenitor understanding. The current simplest interpretation is that SNe~Ic arise
from higher progenitor masses than SNe~Ib.\\
\indent One of the most intriguing results to arise from
SN environmental studies are the properties of the explosion sites of SNe~IIn. These
explosions have generally been proposed to arise from very massive progenitors, possibly
LBVs (see e.g. \citealt{smi08}),
and there is support for this view from e.g. the progenitor detection of SN~2005gl \citep{gal09}.
However, it has been shown that the environments of SNe~IIn (and indeed their close cousins the 
SN `impostors') are somewhat inconsistent with having very high mass progenitors (see above results
and those in \citealt{and12,kel12} and \citealt{hab14}). Indeed, SNe~IIn have very similar environments
as SNe~IIP. In \cite{and12} and \cite{hab14} it has been
argued that this implies that SN~IIn progenitors therefore have similar mass progenitors 
to SNe~IIP, i.e. that they arise from the lower end of the CC SN progenitor mass range.
There is some support for this view from individual SNe~IIn/impostor events.
The transient SN~2008S was constrained to have a progenitor mass of $\sim$10\msun,
from Spitzer pre-explosion images \citep{pri08a}. \cite{smi13} argued that the
Crab nebula is the result of a SN~IIn arising from a low mass progenitor which exploded through the electron
capture SN mechanism, 
and suggested that other SNe~IIn showing a plateau in their light-curves (SNe~IIn-P) may arise
from similar progenitors.\\
\indent One of the issues with analysing statistical samples of SNe~IIn and
`impostors' is their heterogeneous nature (see discussion in \citealt{hab14}).
It is highly probable that there are multiple channels to produce the observed interacting
transient diversity (see e.g. samples published in \citealt{kie12} and
\citealt{tad13}). Further separation of samples into all distinct classes removes
any kind of statistical significance of analysis, and is also hampered by a lack of well observed events.
Such analysis is warranted in the future as the number of well observed events increases.
However, aside from these issues one can still look at the overall sample in a statistical sense, and
as above, this implies that the majority of SNe~IIn do not arise from the most massive stars.\\
\indent Motivated by the above statistical results, \cite{smi15} investigated the environments
of LBV stars within the Milky Way and both the Large and Small Magellanic Clouds (LMC, SMC).
Measuring the distance between LBV stars and the nearest O stars, these authors found
LBVs to be situated in isolated environments, at significant distances from 
collections of O stars. Given that the general assumption is that LBV stars are an evolutionary phase 
which O stars will pass through before becoming WR stars, this is very surprising.
\citeauthor{smi15} argued that this must imply two things. 1) that LBV stars are the mass gainers
within massive binary systems. This would mean that the measured masses for LBV stars are actually 
significantly higher than ZAMS masses. And 2) that the higher mass companions within
these massive binary systems explode first and hence produce a significant kick which displaces
LBV stars from their parent cluster, hence appearing in apparent isolation.\\ 
\indent The location of LBV stars is possibly constraining for 
the nature of both SNe~IIn and SN `impostors'. If these isolated LBV stars are then
the progenitors of a significant fraction of observed SNe~IIn and `impostors', then
\cite{smi15} argue that this could explain the relatively low association of these interacting transients
with host galaxy SF, \textit{without} implying such low progenitor masses as derived for SNe~IIP.
It is also argued that such binary processes will dilute any information that
can be gained on progenitor masses from environment studies. However, while there may indeed
be some dilution of association with \hii\ regions if a large fraction of
SN progenitors have significant velocities, there are several issues
with such claims. Firstly, for an LBV progenitor to obtain a kick velocity, an initial SN 
has to explode. Within the vast majority of massive binary systems the most massive star will
explode first as core evolution is independent of surface mass-loss processes. These first SNe 
are most likely to be of type Ic. 
Hence, in such a system the most massive explosion will be observed to be significantly associated
with their host stellar population, while when the LBV finally evolves
to explosion, it will be less associated because a) it has received
a velocity kick and will hence explode further away from the host \hii\ region, \textit{however
importantly} b) the latter will obviously explode \textit{after} the initial SN,
and hence all previous arguments for an increased association to SF being 
equivalent to younger progenitor ages still apply. Therefore,
the initial mass difference between progenitors within the system will
still be traced by environment properties even in a binary velocity kick scenario. If these
kicked LBVs were the progenitors of SNe~IIn, then the latter would still have lower progenitor
masses than the first SNe to explode, i.e. SNe~Ic, as we have argued above.
The conclusions of \cite{smi15} are intriguing and if true would have significant implications
for stellar life and death. Future observations are needed to test the predictions of their scenario.\\
\indent In conclusion, we believe that differences in the association of SNe
to host galaxy SF indeed imply differences in mean progenitor ages and masses. Exact
progenitor mass constraints are more difficult, especially when tied
in with the complications of binaries above. Finally, we stress that differences
in the association of SNe to SF need to be explained by any valid progenitor scenario.
Evidence for binary progenitors for CC SNe is now outlined in more detail.

\subsubsection{Binary progenitors of CC SNe}
It is now firmly established that the majority of massive stars will have their
evolution significantly affected before explosion because of the influence
of a binary companion \citep{san12_2}.
The presence of a binary companion
may not only affect the evolution of a star, but also affect its positional 
characteristics, i.e. its environment, after the initial star within the system explodes.
Indeed, it has long been suggested that binary evolution can lead to a number 
of observed SNe, especially for the stripped envelope types of Ibc, and IIb (e.g. \citealt{pod92}).\\
\indent \cite{san12_2} found that $\sim$70\%\ of massive O stars will lose mass through
binary interaction (also see \citealt{san13}). This being the case, it is impossible to ignore
the effects of binary interactions on SN progenitors and subsequent SN diversity.
CC SN rates also give strong evidence that at least a fraction of SNe~Ibc
must arise from binaries \citep{kol07,smi11}. \cite{smi11}
analysed the volumetric rates of CC SNe from the Lick Observatory Supernova Search (LOSS),
and compared these to the expected number of SNe within different progenitor mass ranges
assuming a standard IMF. The main conclusion was that single
star scenarios cannot explain the observed rate of SNe~Ibc. These authors created various progenitor
scenarios, and favoured a hybrid scheme where the majority of SNe~Ibc, in particular
the SNe~IIb and SNe~Ib, arise from binary progenitors, but some SNe~Ic are still produced
by single stars. This 
scenario qualitatively agrees with the above findings with respect to environments. Indeed,
\cite{smi11} predicted that following this scenario SNe~Ic should have environments which
are more consistent with higher masses and higher metallicities than SNe~II and SNe~Ib, as is 
concluded in \S\ 3.1 and 3.3. In addition, \cite{eld08} showed that a mix of binary and single star
synthesised progenitor populations most accurately map the observed CC SN ratios and their dependence on metallicity.\\
\indent There is also evidence for binary progenitors through analysis of individual events.
\cite{ber14} modelled the early light-curve of the SN~Ib iPTF13bvn, and constrained its progenitor
to be relatively low mass, which would only be possible in a binary scenario.
In the case of SNe~IIb there are several progenitor detections which indicate a
binary progenitor scenario. After SN~1987A, SN~1993J was the second SN to have a progenitor
detection \citep{ald94}. A companion star was later observed, hence confirming
the binary nature of this explosion \citep{mau04,mau09}. A progenitor
detection was also made for SN~2011dh \citep{van11,mau11}. In this case there 
was some debate as to whether the detection was the actual progenitor, and also 
as to its single or binary nature.
\cite{ber12} argued for a low mass progenitor star, consistent with that detected in
pre-explosion images, which constrained the progenitor to be a binary system.
\cite{fol14_2} have claimed a possible detection of the 
companion star to SN~2011dh in late time Hubble Space Telescope images, which would
be consistent with the above model predictions of \cite{ber12}. In the case
of SNe~Ic we still await a sufficiently nearby and well studied explosion to enable
such constraints.\\
\indent While SNe~II now have a significant number of direct progenitor
detections, there is a lack of pre-explosion SN~Ibc progenitor detections, where only limits have been possible on progenitor luminosities, as shown
by \cite{eld13}. These authors used this evidence, together with the above constraints
on SN relative rates to argue for binary progenitors for the majority of SNe~Ibc, 
and presented binary evolution models to account for observational constraints.\\ 
\indent The discussion in this section clearly shows that there is significant evidence
that binary systems account for at least a fraction of CC SN diversity. Hence, these 
results have to be taken into account when discussing environmental results. However,
while the presence and influence of binary companions may significantly affect the 
nature of the transients themselves, environmental results can still provide 
strong constraints on the nature of progenitors, and further constraints
on the fraction of progenitors that arise from single or binary scenarios.

\subsection{Radial distributions of CC SNe}
The long standing assumption that the centralisation of SNe~Ibc with respect 
to SNe~II within galaxies is due to a progenitor metallicity effect (see e.g.
\citealt{van97,hak09,and09}), can no longer
be deemed valid. We have shown that this result is dominated by SNe within
disturbed galaxies where any metallicity gradients which precede such an assumption
will be much less prevalent, or even non existent. Hence, other explanations
must be explored.
Such discussion has been detailed in \cite{hab10,and11} and \cite{hab12}. We summarise
the main points here.\\
\indent In the central parts of disturbed and interacting systems 
starbursts of violent SF are often frequently found to occur (see e.g. \citealt{lar78,jos85}),
leading to a high turnover of massive stellar birth and death. 
One possibility to explain SNe~Ibc centralisation 
could thus be that we are simply observing systems at an epoch when the 
the most massive stars, i.e. the SNe~Ibc, are exploding following a burst of SF.
However, this would require extreme fine tuning in the sense that the difference
in lifetimes between SNe~II and SNe~Ibc (in any given progenitor scenario) are at most 
several 10s of Myrs. Meanwhile galaxy mergers/interactions have timescales of several 100 Myrs
(see the results of models of merging galaxies in \citealt{mih96}).
Hence, it is not clear how the radial distribution of CC SNe within disturbed galaxies 
can be explained though progenitor age effects. While this could be the case for 
any one given galaxy (e.g. see the SN distribution in Arp 299, \citealt{and11}), 
it is implausible that we are observing the rest of our disturbed sample (such as that displayed in
Fig.\ 8, where it seems obvious that these galaxies display a wide range of starburst
ages) when the dominant the
age of massive stars is only a few Myrs. Again, we encourage the reader to browse
the content of \cite{hab12} for further discussion of this issue, and some more
quantitative arguments.\\
\indent Another possibility is that the binary fraction of massive stars changes with environment.
It is quite possible that the SF within the central parts of disturbed systems is 
dissimilar to that found in `normal' star-forming regions in galaxy disks\footnote{We also
note that it is possible that the binary fraction is higher within the central parts of bright
\hii\ regions, and could be a factor in the interpretation of the results presented in
\S\ 3.1.}.
However, it is not clear in which direction this will go. Denser SF may indeed lead to 
a higher fraction of binaries, however those binaries may be more easily disturbed by 
nearby stellar systems. It is possible that some binary systems are disrupted, but it 
is also possible that those remaining binaries become more compact. The latter 
become close-binaries which are more likely to influence stars before explosion, 
as these hard binaries will lead to mass exchange on the main sequence (e.g. \citealt{dem14}), 
and thus a larger fraction of progenitors could explode
as SNe~Ibc. These issues
are not clear. However, within this discussion there is an important point which is usually ignored.
If a binary system is to produce a SN~Ibc, where the components have insufficient
mass to lose their envelopes through single star processes, then this implies that the companion has \textit{gained}
hydrogen-rich material (or at the very least not lost a significant amount 
of its own envelope). Hence, the second star to explode within the system will 
most likely be hydrogen-rich
and hence a SN~II. Thus, even in a stellar population where all SNe were produced by
progenitors which had undergone binary mass transfer, the
lower mass companion is most likely to explode with a significant fraction
of its hydrogen envelope intact. This would imply that the ratio of SNe~II to SNe~Ibc in such hypothetical
environments would be close to equal. 
We have shown above that in some environments this is clearly not the case
(in the central parts of the extremely disturbed galaxies in the above sample three times more SNe~Ibc than SNe~II have been observed). Hence, while we cannot currently
disprove that SF in the central parts of these galaxies produces a higher fraction of CC SNe
from binaries, it is not clear how this would explain the above results. Nevertheless, 
studies of the fraction of massive stars within binaries as a function
of environment will be an interesting direction for such work in the future.\\
\indent It is also possible that interaction driven SF in the central regions of galaxies
produces a different distribution of stars than those within `normal' SF regions. One way
to explain the centralisation of SNe~Ibc is to assume that these SNe arise from more massive
stars than SNe~II (for which we believe there is significant evidence above), 
and then that the IMF is biased towards higher mass progenitors
in these regions. Such a claim is indeed controversial, as there is significant doubt as to whether
the IMF varies with environment (see \citealt{bas10} for a review). However, there are predictions, 
such as those from \cite{kle07}, who suggest that in extreme SF environments the IMF may indeed be biased
towards high mass stars, with a cut off at lower masses. We do not go into more
detail on this topic here, however we believe that such a possibility needs to be considered when
discussing SN environment properties.

\subsection{Progenitor metallicity constraints}
Single star stellar evolution models predict that there 
should be a significant dependence of the SN~II/SN~Ibc ratio on progenitor metallicity
(e.g. \citealt{heg03,eld04,geo09}). Binary population models (such as \citealt{eld08}) 
predict a much less significant trend with metallicity, as the mass-loss is dominated 
by mass transfer in place of metallicity dependent winds.
Studies of the global properties
of SN host galaxies have observed 
that the SN~II to SN~Ibc ratio is dependent on: galaxy luminosity \citep{pra03,boi09,arc10},
which can be used as a proxy for galaxy metallicity; 
and galaxy metallicity \citep{pri08b}. These studies have argued that these
results imply a significant difference in environment and thus progenitor
metallicities of SNe~II and SNe~Ibc.
However, above we have already shown that the interpretation of the radial
distributions of CC SNe cannot be fully explained through progenitor metallicity effects.
In \S\ 3.3 of this review we have shown that there are only small differences
between CC SN host \hii\ region metallicities between SNe~II and SNe~Ibc. This is then
in conflict with the predictions of single star progenitor scenarios, as 
there is only a \textit{suggestion} that the SN~II to SN~Ibc rate increases
with decreasing oxygen abundance, and only in the lowest \hii\ region bin.
\textit{These results appear to be strong additional evidence that a significant fraction 
of SN~Ibc progenitors arise from binary systems.}\\ 
\indent It is also interesting to note that while 
any differences are statistically marginal, of all the main CC SN types SN~Ic do explode
in the regions with the highest oxygen abundances. In Fig.\ 11 there is some suggestion that
the SN~Ib to SN~Ic rate increases with decreasing metallicity. It was also
shown in \S\ 3.2 that in undisturbed systems, where metallicity gradients are 
measured to be more prevalent, SNe~Ic are indeed found to explode
more centrally than SNe~Ib, lending further weight to a 
progenitor metallicity difference between SNe~Ib and SNe~Ic. Hence, following the 
discussion above in \S\ 4.1, environmental properties appear to 
constrain SNe~Ib to have quite similar progenitor properties in terms of mass and 
metallicity to SNe~II, while SNe~Ic are constrained to have higher mass and metallicity\footnote{Although we note
that the subclass of these events, SNe~Ic BL, appear to favour lower metallicity environments.
See e.g. \cite{mod11}.}
progenitors than both the former. If correct, this hypothesis would imply that
the majority of SNe~Ib arise from binary progenitors, while we suggest that at least a fraction (if not all)
SNe~Ic arise from single star progenitors.

\subsection{Progenitor constraints in the wider context}
In this section we briefly summarise other SN progenitor analyses in the context
of environment work. 
Direct progenitor detections give the most stringent constraints on the origin of SNe.
SNe~IIP are the only type to have a significant number of progenitor detections, where their
initial masses have been constrained to be between 8 to $\sim$20\msun\ (see \citealt{sma09} for a review,
and \citealt{mau13,fra14} for recent examples). This is consistent with environmental constraints
that SNe~IIP arise from the lower end of the CC SN progenitor mass range, and their accurate tracing
of host galaxy near-UV emission (Fig.\ 2). There are suggestions from environmental
studies that SNe~IIL have higher masses than SNe~IIP (see e.g. Fig.\ 22). Only a couple
of progenitor detections of SNe~IIL exist, however they do appear to be higher in mass than 
SNe~IIP \citep{eli10,fra10,eli11}. The lack of detections of SN~Ibc progenitors, as shown in
\cite{eld13}, has been used to argue for binary progenitors for these SNe. However, we stress that any such
study should separate SNe~Ib from SNe~Ic given their distinct environments.\\
\indent \cite{lel10} investigated the spatial distribution of WR stars within
galaxies. Using a similar technique to the pixel statistics methods outlined above,
these authors found that WR stars are found to be located in similar environments
as SNe~Ibc, thus supporting the possibility that some of the progenitors of the latter
are high mass single stars. It was also found that SNe~Ic showed environmental properties
more similar to WC stars, and SNe~Ib to WN stars, consistent with the scenario that
SNe~Ic are explosions of stars that have been more heavily stripped of their envelopes.
It may be interesting to further these studies for other stellar types, such as red
and blue supergiants, and look in more detail at the association of different stars to
more direct tracers of SF, such as \ha. We also recall the investigation
of \cite{smi15}, which sheds light on the spatial distribution of different massive stars
(and hence possible SN progenitors) with respect 
to each other.\\
\indent There has also been work constraining SN progenitors through analysis of resolved stellar
populations within very nearby galaxies \citep{gog09_2,mur11,wil14,jen14}. These studies
estimate the age of resolved parent stellar populations (on scales of $\sim$50 pc) and translate these to progenitor masses.
As we observe additional SNe in the very nearby Universe, and as the next generation of extremely
large telescopes come on line, the possibilities of these studies will only increase.\\
\indent \cite{lym14_2} studied
the properties of a sample of 38 stripped envelope SNe (SNe~IIb, Ib, Ic, Ic-BL). These authors
concluded that derived ejecta masses imply that low mass binary progenitors
dominate the progenitor channel of these SNe. Interestingly, they did find that SNe~Ic have
higher ejecta masses than SNe~Ib, perhaps implying higher mass progenitors for the latter, as
suggested by environmental results above.

\subsection{Selection effects of environment observations and SN samples}
There are now a large number of SN environmental studies, as has been discussed
throughout this review. There are many selection biases that may affect 
different studies, both in the nature of the environment studies, and 
in the nature of the SN samples themselves. Here we briefly summarise these, but we
refer the reader to the literature cited throughout this review for more complete
discussions.\\
\indent Exact objective SN typing is still an issue today when one has insufficient light-curve
or spectral information. Indeed, many SNe are simply classified spectroscopically and 
then are not followed any further. This problem may be
more important for analyses based on old literature SNe, when less data were obtained, 
and less was known
about the diversity of SNe and its physical origin. However, its is not
clear why this problem should bias environment results in any particular direction.\\
\indent In \S\ 3.3 we combined the measurements of many different host \hii\ region oxygen abundance
studies. While these studies were all of resolved galaxies, the redshift
range of their samples is quite different, with some very nearby studies such as \cite{kun13,kun13_2}
combined with more distant samples. As one goes to larger distance the reduction in
spatial resolution means that one's measurement is over a much larger environment.
Other than diluting the information of the exact parent stellar population, it is unclear
how this will affect the results of environment studies. One issue that is often 
cited is that if there is any significant difference in the luminosities of different SNe
which are being compared, then this will result in a bias against the detection of one SN type 
over the other against e.g. bright \hii\ regions, the brighter centres of host galaxies, 
or within regions of higher host galaxy extinction. While SNe~Ia are significantly brighter
than CC SNe, there is actually little difference between the peak absolute magnitudes
of SNe~II and SNe~Ibc, as shown by SNe discovered within the LOSS survey \citep{li11}.
Hence, \textit{there is little evidence for any selection bias that could explain the preference
of SNe~Ibc for exploding in both more central regions of host galaxies, and within bright \hii\ regions.}\\
\indent The exact make-up of host galaxy samples is also often 
discussed in terms of a bias between different studies. While traditional 
SN searches have concentrated on targeting massive, bright galaxies, there now exist 
a number of un-targeted searches which scan the sky searching for SNe, and find many more SNe in low
surface brightness galaxies than the former. Undertaking environment studies in these different
samples will affect the relative numbers of studied SNe and may also change the nature of those
SN populations. However, it is interesting to note (as already stated above) that \cite{sto13}
found that a SN~II sample from an un-targeted search had a very similar \hii\ region oxygen abundance
distribution as that that derived from a sample of literature SNe discovered by
targeted searches.\\
\indent All of the above issues need to be taken into account when studying SN environments,
and future studies will strive to reduce any systematic biases in their samples, to 
obtain the most robust progenitor and transient constraints.

\section{Current and future prospects}
SN environment studies are still in their infancy, and current/future instrumentation
promises to further open the field for new results and analysis. There are two specific
avenues which we believe will be pursued by environment studies to significantly increase
insight on SN progenitors. The first involves
the actual transients. This advancement was introduced in \S\ 3.7, and involves having 
large samples of well characterised SNe. From these samples one can then tie specific environment
properties to specific SN features which may be more physical in nature than simple type classifications.
The second involves new observations enabled by current and future instrumentation, and specifically the
advantages gained for environment studies by using IFU observations. We now discuss this
in more detail, and show some preliminary analysis in that direction.\\

\begin{figure*}
\begin{center}
\includegraphics[width=17cm]{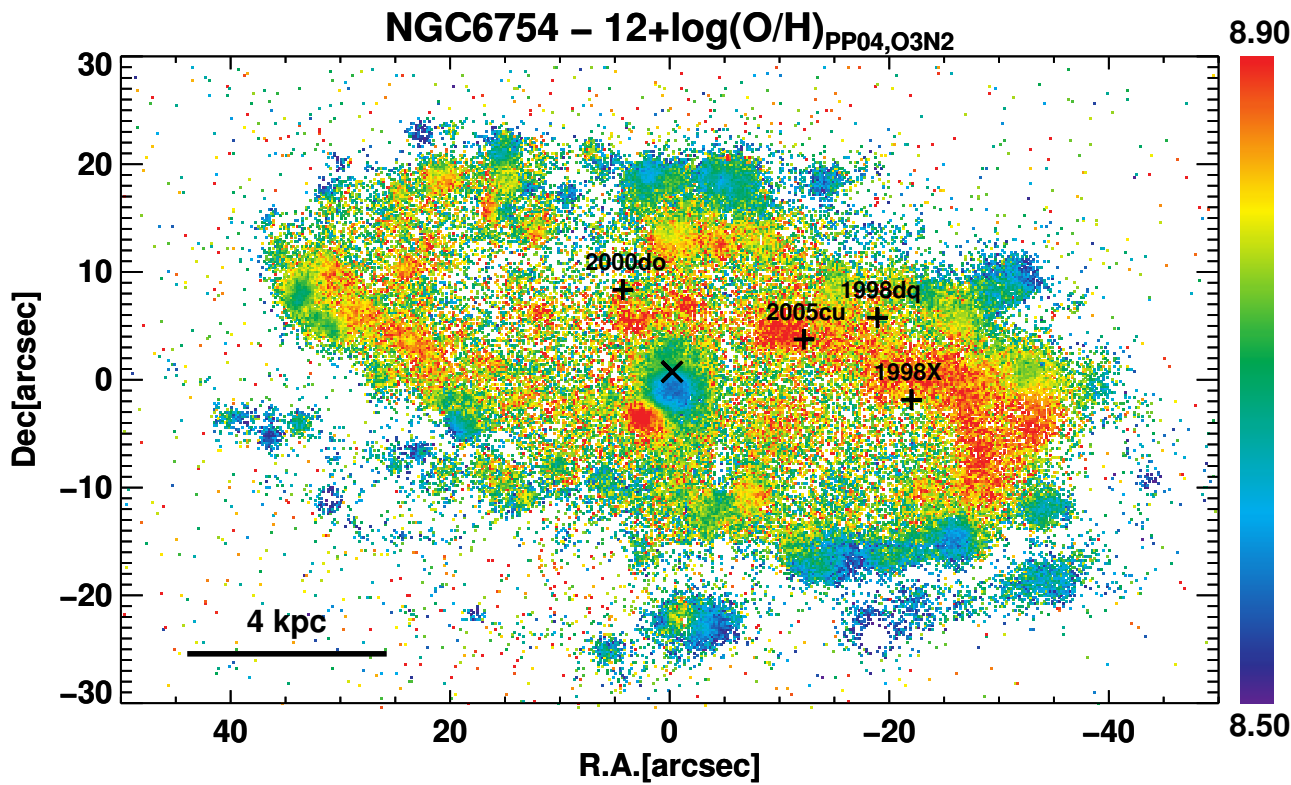}
\caption{PP04 O3N2 metallicity map of NGC6754, with SN explosion sites indicated.}
\end{center}
\end{figure*}

\subsection{New types of environment observations}
We have already commented on the central importance
of wide-field IFUs for future studies of SN host galaxies.
This enables simultaneous spatial
and spectral information over the entire extent of host galaxies.
Indeed, the usefulness of IFU observations for SN host environment studies
has already been demonstrated by e.g. \cite{sta12,rig13,kun13,kun13_2} and
\cite{gal14}.\\ 
\indent MUSE mounted on the VLT telescope promises to 
significantly enhance the capabilities of IFU instrumentation for environment studies,
having a FOV of 1\am$\times$1\am, extremely good spatial resolution (0.2\as\ spaxel size), and the power
of being attached to an 8m class telescope.
We recently obtained Science Verification data (P.I. Galbany), to observe
the galaxy NGC 6754 which has been host to four SNe. A full analysis of that galaxy and 
other data taken during SV in terms of SN environmental parameters will be
presented in Galbany et al. (in preparation). Here we show a glimpse of the data
obtained and show what can be achieved through such analysis. (Note, an analysis
of the \hii\ region population of NGC 6754 using these data has been presented
in \citealt{san15}.)
In Fig.\ 25 we present the spatial metallicity map of NGC 6754 on the PP04 O3N2 scale,
with SN positions indicated. In Fig.\ 26 a histogram of all measured metallicities
is shown, with values estimated at exact SN sites indicated. These figures
show the power of these data as compared to previous work, much of which was discussed above, where
in the latter either spatial information of a specific emission line (e.g. \ha) has been analysed,
or spectral information but in one small scale region. With current state-of-the-art
instrumentation such as MUSE one can obtain all these parameters with one single telescope pointing.\\
\indent Additional instruments are also coming on-line to further these IFU possibilities, such
as SAMI \citep{bry12}, and MANGA \citep{law14}. WiFeS \citep{dop10,chi14} is another instrument
which is being used for these types of studies.
These multi-IFU instruments will observe thousands
of galaxies. While SN environment science may not initially be the key science drivers, when one observes
such quantities of galaxies, then one naturally obtains statistically significant samples of SN host 
galaxies, which can be used for environment studies. Using these instruments one can essentially
analyse host galaxies/environments in the manner described in \S\ 2.1, 2.2 and 2.3 \textit{simultaneously}, i.e.
one can extract any emission line (obviously within the constraints of any given instrument), or
broad-band `colour' and proceed with `spaxel statistics'. This will enable environment parameters such
as metallicity to not only be compared between SN types, but also between SN environments \textit{and all
other environments within their host galaxy} (this in essence is what is done above for the \ha\ pixel
statistics). This is what is shown in Fig.\ 26: a histogram of all host galaxy environment metallicity
measurements.
In this particular galaxy it is observed that the SNe~Ia explode within the peak of the metallicity distribution,
while the SNe~II appear to explode in regions of higher oxygen abundance. We note here that this analysis
is very preliminary, and is shown here to present future possibilities. Published values
may change once a more thorough analysis is achieved.\\
\indent While large scale statistical environment analysis will hugely benefit with the 
advent of wide-field IFU instrumentation, it is also interesting to attempt to observe environments
with the best spatial resolution in order to observe individual progenitor parent stellar clusters.
In order to obtain the best spatial resolution, adaptive optics (AO) helps to overcome the limitations
of atmospheric conditions (i.e. seeing) during an observation. In this fashion we have
obtained data of a handful of SN environments with SINFONI \citep{eis03}, a near-IR
IFU instrument. In Fig.\ 27 we present an example of the gain in spatial resolution using
near-IR AO observations, where SINFONI observations are compared to those of an optical IFU of the
same environment. This shows that what appears to be a single cluster in an optical
observation is resolved into several clusters in SINFONI observations. In the near-IR
one can use the presence or absence of Br-gamma emission to constrain the age of the 
environment/parent cluster (more details will be presented in Kuncarayakti et al. in preparation). 
Better spatial resolution also enables us to alleviate SF history issues.\\
\indent The above examples show how the field of environments is set to evolve over the 
coming years. In addition, one assumes that investigators will find other innovative
ideas to use other instruments to study SN environments. In combination 
with growing databases of high quality follow-up data, the coming years
promise to be particularly fruitful for the environment and SN fields.

\begin{figure}
\begin{center}
\includegraphics[width=\columnwidth]{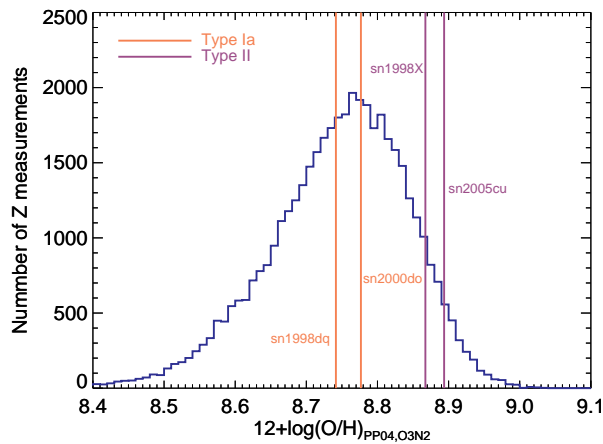}
\caption{Histogram of PP04 O3N2 metallicity measurements of NGC 6754, with SN environment
metallicity positions indicated.}
\end{center}
\end{figure}

\begin{figure}
\begin{center}
\includegraphics[width=\columnwidth]{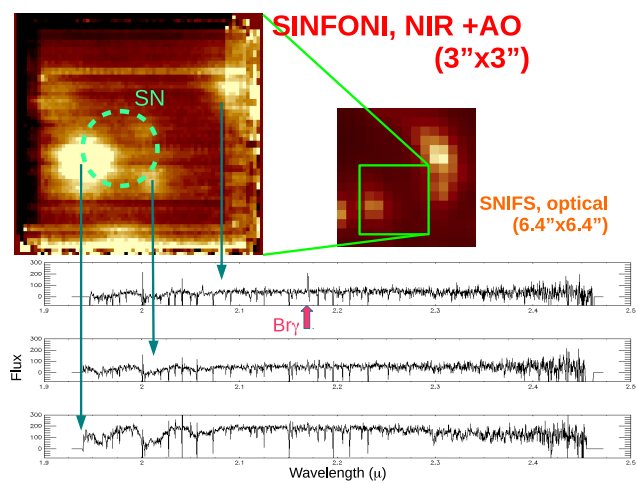}
\caption{Example of the advantages of moving to the near-IR where one can obtain much higher
spatial resolution due to adaptive optics corrections. A comparison is made between optical IFU 
(SNIFS)
observations, and those obtained with SINFONI of the same SN environment. Also shown are the extracted
near-IR spectra, where the Br-gamma emission line is indicated.
(A similar figure is also shown in \citealt{kun14}.)}
\end{center}
\end{figure}

\section{Summary and conclusions}
The field of SN environments is still relatively young, but is growing into an important
area of investigation. Major differences in the environmental properties of different
SNe exist, and any hypothesised progenitor scenario must be able to explain such trends.\\
\indent In this manuscript we have reviewed the current status of SN environment studies,
and how these can be used to constrain SN progenitors. Differences in the environments
of SNe~II, SNe~Ib and SNe~Ic appear to imply that the progenitor mass and metallicity ranges of
the first two are quite similar, with larger values for the latter.
This is explained if the majority of SNe~Ib arise from binary systems, however we note 
that environmental results still allow for a significant number of SNe~Ic to arise
from higher mass, and possibly single star progenitors. These conclusions are
in qualitative agreement with contemporary studies of SNe. 
The environmental properties of SNe~IIn test our understanding of these events and their progenitors,
and it will be intriguing to see what further investigation brings to this debate. 
While studies of SN~Ia environments are perhaps less constraining, there
indeed exist interesting results.\\
\indent SN environment studies are rapidly evolving with new instrumentation, and 
ever larger SN samples to study. Future years promise to be rich 
with new analysis, and further constraints on the large diversity of SNe.

\begin{acknowledgements}
In no particular order we thank: Mario Hamuy, Francisco F\"orster,
Santiago Gonz\'alez, Stephen Smartt, Nathan Smith, Maryam Modjaz,
Gaston Folatelli, Melina Bersten, Mark Phillips, Max Stritzinger,
Paul Crowther, Giuliano Pignata, Franz Bauer, Morgan Fraser, Joe Lyman
for many fruitful discussions which contributed to the content
of this review.
Support for HK and LG is provided by the Ministry of 
Economy, Development, and Tourism's Millennium Science 
Initiative through grant IC120009, awarded to The Millennium 
Institute of Astrophysics, MAS. HK acknowledges support by 
CONICYT through FONDECYT grant 3140563, and LG through grant 3140566.\\
\end{acknowledgements}

\bibliographystyle{apj}
\bibliography{Reference}

\end{document}